\documentclass{article}

\usepackage{PRIMEarxiv}
\usepackage[utf8]{inputenc} 
\usepackage[T1]{fontenc}    
\usepackage{hyperref}       
\usepackage{url}            
\usepackage{booktabs}       
\usepackage{fancyhdr}       
\usepackage{graphicx}       
\usepackage{multirow}
\usepackage{subfigure}
\usepackage{colortbl}
\usepackage{subfiles}
\usepackage{algorithm}
\usepackage{algorithmic}
\usepackage{xcolor}
\usepackage{amsmath}
\usepackage{amssymb}
\usepackage{natbib}
\usepackage{subfiles}

\title{ A Lightweight Framework for Adaptive Retrieval in Code Completion with Critique Model}

\author{
  Wenrui Zhang, Tiehang Fu, Ting Yuan, Ge Zhang, Dong Chen$^*$, Jie Wang\thanks{Corresponding Authors}\\
  \\
  Huawei Technologies Co., Ltd \\
  \texttt{zhangwenrui@u.nus.edu} \\
  \texttt{2018fth@whu.edu.cn} \\
  \texttt{\{yuanting811,zhangge30\}}@huawei.com\\
  \texttt{jameschennerd@gmail.com}\\
  \texttt{jiewmurphy@163.com}
}

\newcommand{\ourmethod}{CARD}
\newcommand{\ourmodule}{Estimator}
\newcommand{\benchmark}{RepoEval-M}

\newcommand{\RG}{\textit{RG} }

\begin{document}
\maketitle

\begin{abstract}
Recent advancements in Retrieval-Augmented Generation have significantly enhanced code completion at the repository level. Various RAG-based code completion systems are proposed based on different design choices. For instance, gaining more effectiveness at the cost of repeating the retrieval-generation process multiple times. However, the indiscriminate use of retrieval in current methods reveals issues pertaining to both efficiency and effectiveness, as a considerable portion of retrievals are unnecessary and may introduce unhelpful or even harmful suggestions to code language models.

To address these challenges, we introduce \ourmethod, a lightweight critique method designed to provide insights into the necessity of retrievals and select the optimal answer from multiple predictions. \text{\ourmethod} can seamlessly integrate into any RAG-based code completion system. Our evaluation shows that \text{\ourmethod} saves 21\% to 46\% times of retrieval for Line completion, 14\% to 40\% times of retrieval for API completion, and 6\% to 46.5\% times of retrieval for function completion respectively, while improving the accuracy. \text{\ourmethod} reduces latency ranging from 16\% to 83\%. \text{\ourmethod} is generalizable to different LMs, retrievers, and programming languages. It is lightweight with training in few seconds and inference in few milliseconds.

\end{abstract}

\section{Introduction}
Language models (LMs) for code trained on massive source code have established a new state-of-the-art in code completion \cite{rozière2024code,lozhkov2024starcoder,guo2024deepseekcoder}.
However, LMs often suffer from "hallucination", which leads to the fabrication of what can be referred to in a repository.
The technique Retrieval-Augmented Generation (RAG) alleviates the ``hallucination".
It provides dynamic access to the latest external data and transparency in response generation without time-consuming and computation-extensive fine-tuning. 
In practice, various RAG-based code completion systems are proposed based on different design choices~\cite{lu2022reacc,clement2021longrange,liang2024repofuse,cheng2024dataflowguided,eghbali2024dehallucinator,tan2024promptbased,zhang2023repocoder,shao2023enhancing}: what to retrieve, how to assemble, how many iterations (iterative RAG), and etc.

\begin{figure}
    \centering
    \subfigure[The distribution of Edit Similarity (ES). Each bucket represents the ratio of ES values falling in a range of 0.1.]{
    \label{subfig:es}
    \includegraphics[width=0.7\linewidth]{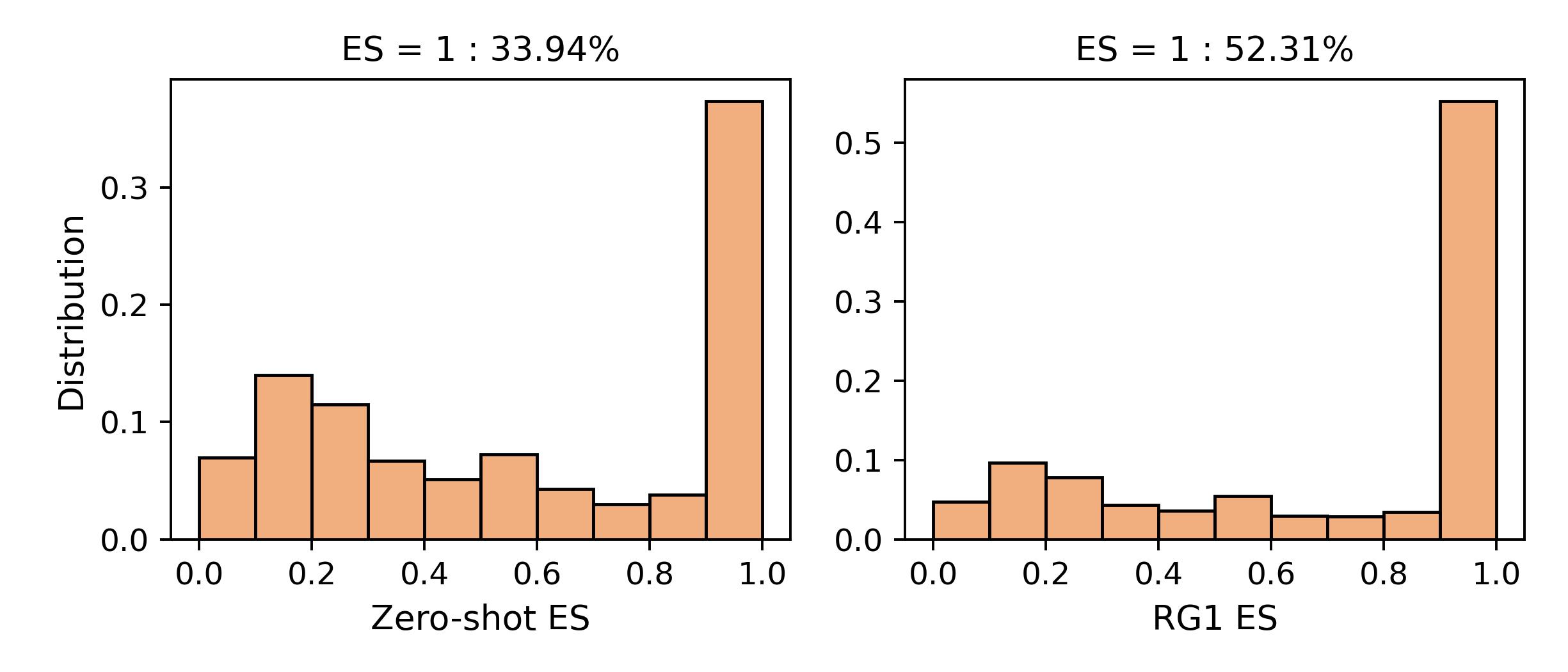}
    }
    
    \subfigure[The cumulative distribution of ES improvement.]{
    \label{subfig:esdiff}
    \includegraphics[width=0.7\linewidth]{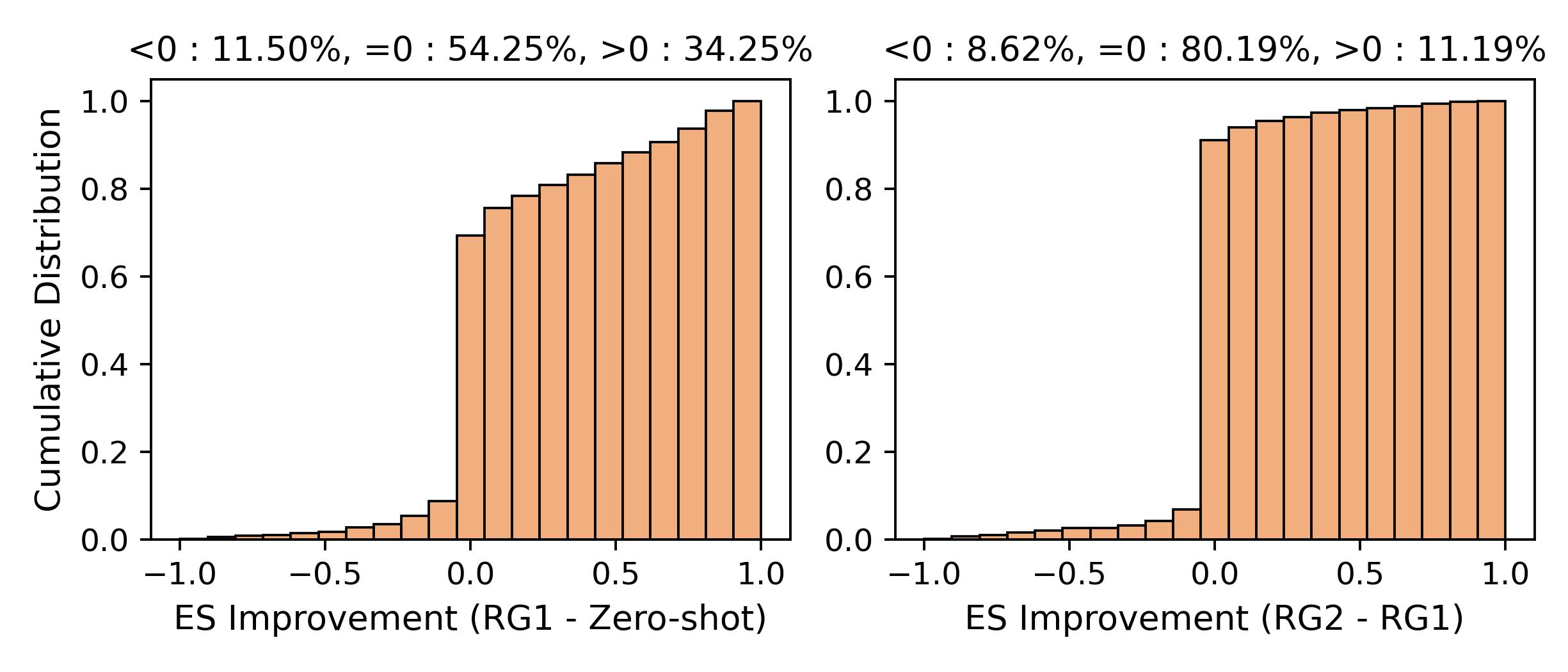}
    }
    \caption{The distribution of ES (top) and improvement of ES (bottom) on the line dataset of RepoEval \cite{zhang2023repocoder} considering zero-shot setup (left) and iterative RAG (right) respectively. RG$i$ stands for the $i$-th Retrieval-Generation. }
    \label{fig:esdiff_dist}
\end{figure}

\begin{figure*}
    \centering
    \includegraphics[width=\linewidth]{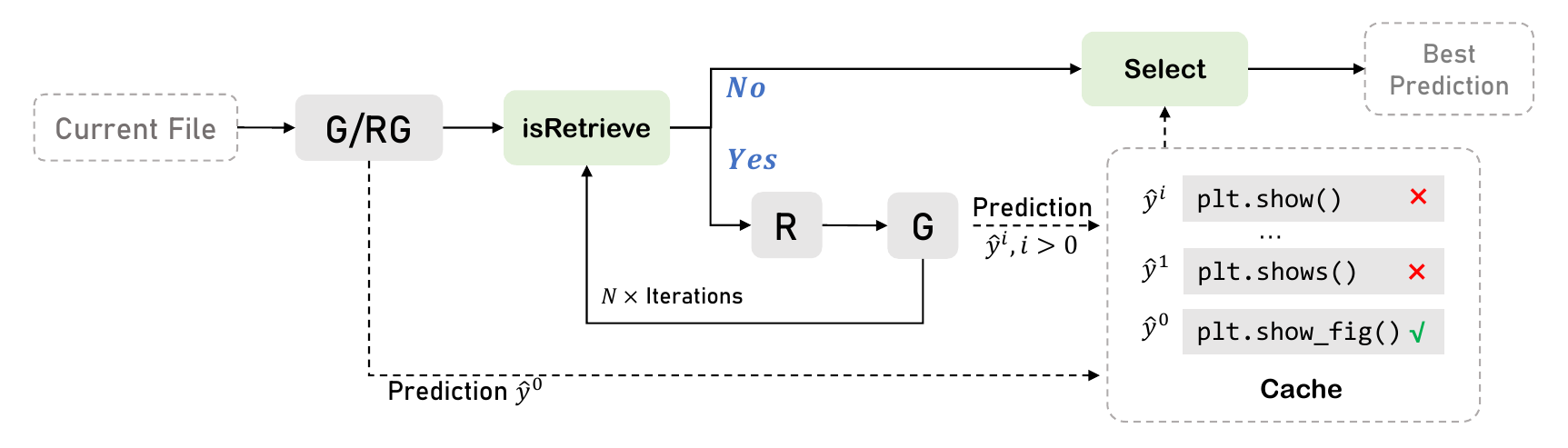}
    \caption{CARD: Suppose that a RAG-based code completion system uses the function $R$ for retrieval and uses the function $G$ for generation. Whenever a code prediction is generated with or without retrieved information, the function $isRetrieve$ can be queried to determine whether retrieval is necessary. When multiple predictions are generated from different iterations, the function $select$ can be queried to determine which prediction is the best. These two functions introduce little overhead and can be used independently.
}
    \label{fig:intro_overflow}
\end{figure*}

Despite the different design sensibilities in a RAG-based system, a crucial problem is: \emph{Is the retrieval and augmentation beneficial?}  
To understand the limits and opportunities of adopting RAG, we perform a study on the line dataset of RepoEval~\cite{zhang2023repocoder} using CodeLlama 7B \cite{rozière2024code}. As the Figure~\ref{fig:esdiff_dist} shows, we find that
(a) 33.9\% of the predictions directly generated by LM is correct. This number goes much larger for a more powerful LM. The retrieval should be avoided under such good circumstances.
(b) about 10\% of generations will degenerate through \RG (denoted as the \textit{degenerated cases} problem), where the deteriorated generation should be rejected.
It reveals that invariant retrieval (always retrieve) can introduce irrelevant or harmful information to code LMs, impacting both efficiency and effectiveness.

To address these challenges, we provide a lightweight method \text{\ourmethod}, based on a machine learning model estimating the uncertainty of LMs, to conduct adaptive and selective RAG for code as shown in Figure~\ref{fig:intro_overflow}.
It exposes two functions tackling the above two problems respectively:
(1) the function \textbf{\textit{isRetrieve}} decides whether a retrieval is required, avoiding unnecessary cost.
(2) the function \textbf{\textit{Select}} returns the best generation among all candidate predictions.

The design of \text{\ourmethod} reflects three core strengths: 
\begin{enumerate}
    \item \textbf{Easy to deploy.} A lightweight plugin design can be seamlessly integrated into any RAG system. It does not require any training or changes for the generator or retriever.
    \item \textbf{Reduce non-beneficial retrievals.} With uncertainty estimation, the \textbf{\textit{isRetrieve}} function will stop iterating when the current generation is ``good" enough.
    \item \textbf{Reduce degeneration.} The selective mechanism mitigates the problem of \textit{degenerated cases} in constant time.
\end{enumerate}
In summary, the main contributions of this paper are:
\begin{itemize}
    \item We present a novel lightweight approach \text{\ourmethod}, which enhances the effectiveness and efficiency of any RAG system for code completion.  
    \item We present a multi-language benchmark \text{\benchmark} containing a total of 4k samples from 8 popular repositories, to evaluate the repository-level line completion task. The programming languages include C, Python, Java, and JavaScript. \text{\benchmark} is available at \url{https://github.com/xxx} (anonymous for review).
    \item   We evaluate \text{\ourmethod} on RepoEval \cite{zhang2023repocoder} and \text{\benchmark}. The result shows that \text{\ourmethod} saves 21\% to 46\% times of RG for Line completion, 14\% to 40\% times of RG for API completion, and 6\% to 46.5\% times of RG for function completion respectively, while improving the accuracy.
    In the meanwhile, \text{\ourmethod} reduces latency ranging from 16\% to 83\%. 
    \text{\ourmethod} is generalizable to different LMs, retrievers, and programming languages. It is lightweight with training in few seconds and inference in few milliseconds.
\end{itemize}

\section{Methodology}
\label{sec:method}

As Figure~\ref{fig:intro_overflow} shows, the RAG-based system \text{\ourmethod} targets at contains the following main functions:
\begin{itemize}
 \item  \textbf{\textit{R}} retrieves relevant information which is included in the prompt for later generation.
\item \textbf{\textit{G}} utilizes a LM $\mathcal{M}_\theta$ to generate code with the input prompt $X$, and returns a pair $(\mathbf{\hat{y}},\hat{logits})$ containing predicted token sequence $\mathbf{\hat{y}}$ and its corresponding logits $\hat{logits}$. For multiple generation instances of one code completion task, we use the superscript to differentiate them, i.e.,   $X^i$, $\mathbf{\hat{y}^i}$, and $\hat{logits}^i$.

\end{itemize}

Our method \text{\ourmethod} provides two critique functions for improving the efficiency and effectiveness of the RAG-based system.
\begin{itemize}
\item \textbf{\textit{isRetrieve}} that follows the $G$ action and uses the generated pair $(\mathbf{\hat{y}^i},\hat{logits}^i)$ to evaluate if retrieval is necessary. 
\item \textbf{\textit{Select}} that aims to choose the best prediction among $\mathbf{\hat{y}^0}\sim\mathbf{\hat{y}^i}$ when no more retrieval is needed (Section \ref{sec:selective_acc}).

\end{itemize}

Our $isRetrieve$ and $Select$ use an uncertainty estimation model denoted as \textit{\ourmodule} for evaluating the quality of a generated prediction, and further make the above suggestions. We introduce them respectively in the following sections. 
\begin{table}[ht]
\centering
\caption{Operators to construct features. All operators are applied on a list of values $\mathbf{x}$, whose length is $N$.}
\label{tab:operators}
\begin{tabular}{c|c}
\toprule
Operation    & Feature ($\mathbf{x}$ = $\{p_t(y_t)\}$ or $\mathbf{x} = \{H_t\}$) \\
\midrule
Max     & $\max_{x\in\mathbf{x}}(x)$     \\
Min     & $\min_{x\in\mathbf{x}}(x)$     \\
Avg     & $\sum_{x\in\mathbf{x}}(x) / N $   \\
Standard Deviation& $\sqrt{\sum_{x\in\mathbf{x}}(x - Avg(\mathbf{x}))^2 / N}$          \\
Product & $\prod_{x\in\mathbf{x}}(x)$          \\
Geometric Avg& $Product(\mathbf{x})^{\frac{1}{N}}$          \\
Len     & N          \\
\bottomrule
\end{tabular}
\end{table}

\subsection{Uncertainty estimation} \label{sec:aider}
\textit{\ourmodule} aims at scoring the predictions of an LM, i.e., pair of $(\mathbf{\hat{y}},\hat{logits})$. 
We leverage the idea of uncertainty estimation introduced by \citet{liu2024uncertainty}. 
We formulate a supervised regression task that fits a model to the dataset $\mathcal{D}=\{(\mathbf{z}, s)\}$. 
$\mathbf{z}$ is a hand-crafted statistical feature vector of $\hat{logits}$ and $s$ is the target score of the prediction regard to the LM. 
We introduce how to construct the dataset $\mathcal{D}$ and train the model as follows. 

\textbf{Dataset construction.}
$\mathbf{z}$ is a feature vector summarizing information of entropy and probability for the predictions, where probability estimates token-level uncertainty~\cite{gupta2024language} and entropy measures the degree of chaos of the predictions~\cite{10172803}. 
Formally, for any candidate token $v$ at the time step $t$, the probability of $v$ is denoted as $p_t(v)$ and obtained by 
 $$p_t(v) = p(v|X,\hat{y}_{1},\hat{y}_{2},\cdots,\hat{y}_{t-1})=\frac{\exp(\hat{logits}_t(v))}{\sum_{v'\in\mathcal{V}}\exp(\hat{logits}_t(v'))}$$ 
where $\mathbf{\hat{y}}_t$ and $\hat{logits}_t(v)$ is the output token and output logits of token $v$ at time step $t$ respectively, and $\mathcal{V}$ is the complete vocabulary of $\mathcal{M}_\theta$. 
Considering the probability distribution of each token, the information entropy at time step $t$ is denoted as $H_t$ and obtained following 
\begin{align*}
    H_t = H(p(\cdot|X,\hat{y}_{1},\hat{y}_{2},\cdots,\hat{y}_{t-1})) =-\sum_{v\in\mathcal{V}}p_t(v)\log(p_t(v))
\end{align*}

To obtain $\mathbf{z}$, we apply seven commonly used statistical operators on $\{p_t(\hat{y}_t)\}$ and $\{H_t\}, t=1,2,\cdots,N$ (listed in Table \ref{tab:operators}).
They form a feature vector with the length of 13. The $Len$ operator returns the same values for entropy and probability.

To obtain $s$, we apply the function edit similarity (ES) for each $\hat{y}$ as follows. $$ES(\mathbf{y},\mathbf{\hat{y}})=1-\frac{\text{Lev}(\mathbf{y}, \mathbf{\hat{y}})}{\max(|\mathbf{y}|,|\mathbf{\hat{y}}|)}$$ where $\text{Lev}$ is the Levenshtein distance~\cite{Levenshtein1965BinaryCC}. 

\textbf{Training within seconds}
We fit a decision tree-based model LightGBM ~\cite{NIPS2017_6449f44a} to dataset $\mathcal{D}$, and denote the trained model as \textit{\ourmodule}. 
LightGBM is designed especially for high training and inference speed (training in few seconds and inference in few milliseconds in our experiment) and low memory consumption. 
In practice, \textit{\ourmodule} can be substituted by other machine-learning regression models.

\begin{figure}
    \centering
    \includegraphics[width=0.8\linewidth]{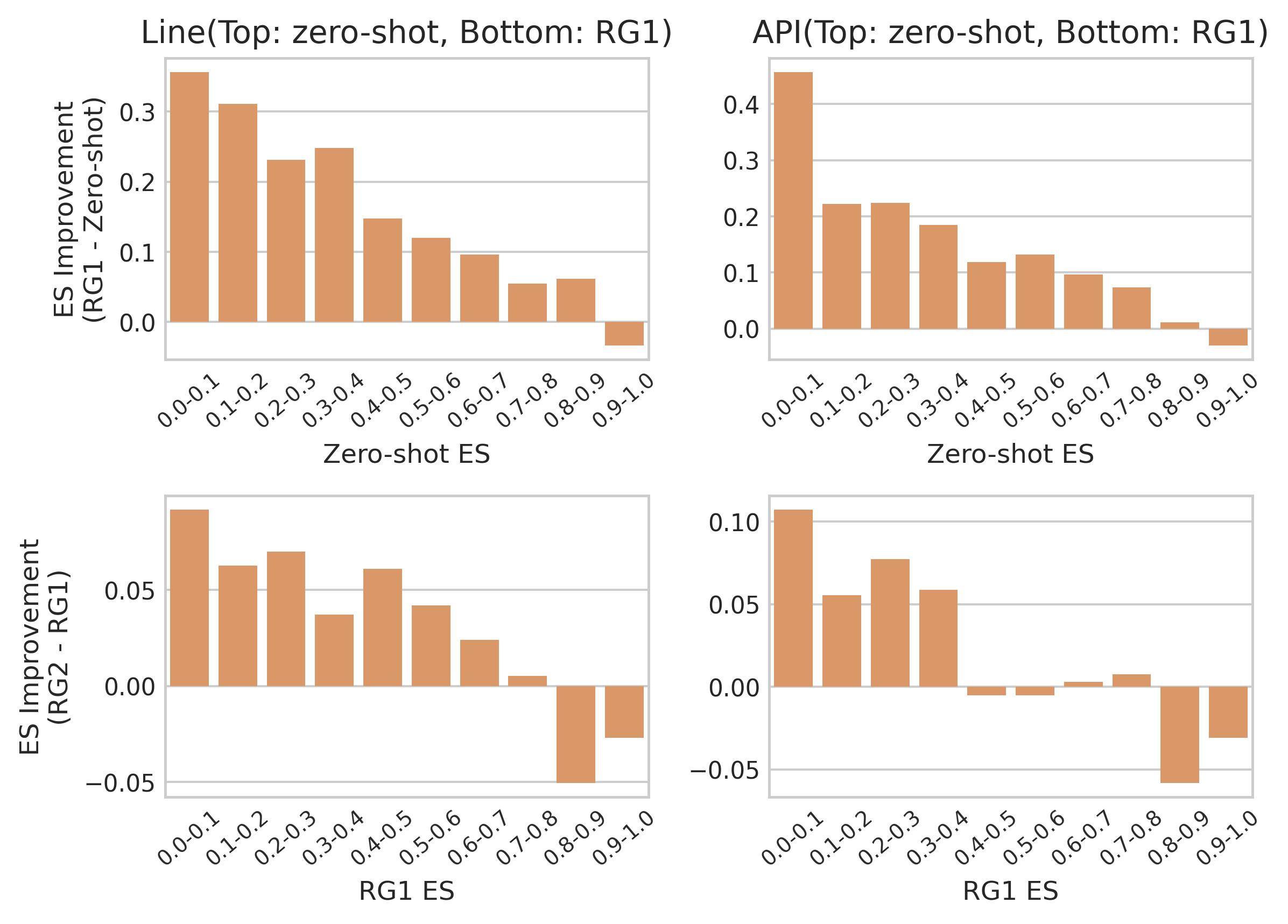}
    \caption{Improvement of RAG on two datasets (Left: line, Right: API) of RepoEval~\cite{zhang2023repocoder} with CodeLlama 7b as code LM. The x-axis represents the ES of zero-shot generation (top) and RG1 generation (bottom), and the y-axis represents the improvement of ES via RAG.}
    \label{fig:esdiff}
\end{figure}

\subsection{Adaptive retrieval} \label{sec:adaptive_rag}
We expose the function $isRetrieve$ to give suggestions about whether retrieval is needed according to the score $\hat{s}$ of the current generated code prediction estimated by \textit{\ourmodule}. 
We decide to conduct or continue RAG only when the predicted $\hat{s}$ is less than the given threshold $T_{RAG}\in[0,1]$. Formally, the function deciding whether to retrieve based on the current logits $\hat{logits}$ is denoted as $isRetrieve(\hat{logits}, T_{RAG})$, following:
\begin{equation*}
    isRetrieve(\hat{logits}, T_{RAG})=\begin{cases}
        \textit{true}, &\hat{s}<T_{RAG}\\
        \textit{false}, &\hat{s}\geq T_{RAG}\\
    \end{cases}\\
\end{equation*}, where $\hat{s} = \textit{\ourmodule}(\hat{logits})$.
The higher $T_{RAG}$ is, the more retrieval will be conducted.

\subsection{Selective accept} \label{sec:selective_acc}
Facing the problem of \textit{degenerated cases} caused by ineffective retrieval, we expose the function $Select$ to determine the best one among multiple code predictions.

Intuitively, the generation with a higher predicted score $\hat{s}$ should be preserved. Formally, for the a predicted pair $(\mathbf{\hat{y}^i},\hat{logits}^i)$ and another predicted pair $(\mathbf{\hat{y}^j},\hat{logits}^j)$ (supposing $j>i$), the $Select$ function is:
\begin{equation*}
    Select(\hat{logits}^{i}, \hat{logits}^{j}, T_{ACC})=\begin{cases}
        \textit{ture}  &, \frac{\hat{s}^{j}}{\hat{s}^{i} + \epsilon}<T_{ACC}\\
        \textit{false} &, \frac{\hat{s}^{j}}{\hat{s}^{i} + \epsilon}\geq T_{ACC}
    \end{cases}
\end{equation*}
where $T_{ACC}$ is a given threshold and could be changed in different iterations, and $\epsilon$ is a small value to prevent the denominator from $0$. When the $Select$ function returns $true$, the generation $\mathbf{\hat{y}^{i}}$ is kept, and vice versa. The threshold $T_{ACC}$ can be adjusted by the users, where the higher $T_{ACC}$ is, the more preference is attached to the generation $\mathbf{\hat{y^i}}$.

\subsection{Applications in RAG-based code completion.}\label{sec:application}
\text{\ourmethod} can be flexibly integrated into any RAG-based code completion system with any prompt setup, retriever, or generator. 
$isRetrieve$ follows a $G$ process, and $Select$ can be used when there are multiple predictions. They are independent and can be used separately.

For instance, we describe how \text{\ourmethod} gains benefits in two common RAG designs (shown in Figure \ref{fig:intro_overflow}).  
In the following text, we specify $(\mathbf{\hat{y}^i}, \hat{logits}^i)$ as the generation after $i$ iterations of RG, and zero-shot generation is regarded as the $0$-th iteration.

\textbf{Single RAG}.
In this scenario, $G$ is triggered first with the prediction result $\mathbf{\hat{y}^0}$ and $\hat{logits}^0$. Then $isRetrieve$ is queried to determine whether needs to retrieve. If the decision is not, the $\mathbf{\hat{y}^0}$ is sent to $Select$ for deciding the final answer. Otherwise, a retrieval is performed and a new prediction $\mathbf{\hat{y}^1}$ is generated. These two predictions are sent to $Select$ to decide the best prediction. 

\textbf{Iterative RAG}. 
Iterative RAG begins with an retrieval and generate process. $isRetrieve$ is used to determine if another iteration is needed. The iteration will stop if the decision is not. $Select$ is used to choose the best prediction among all iterations. In this case, different iterations bring varying performance gains, implying that the need for retrieval and our preference for the outputs change across iterations. Thus, the $T_{RAG}$ and $T_{ACC}$ can be adjusted separately for different iterations.

We determine the setups for these two thresholds based on prior knowledge and preference according to statistical distributions of ES values among different iterations. For instance, as iteration goes on, the need for retrieval drops because the remaining wrong predictions are less likely to correct, and $T_{RAG}$ should decrease.  In addition, the performance gain declines as more RGs are conducted, and $T_{ACC}$ can be set larger for later iterations to accept more former generations.

$T_{ACC}$ is recommended to be close to 1 (equal to directly comparing two $\hat{s}$), and the $T_{RAG}$ can be flexibly adjusted to balance the latency and precision. 

The whole process of \text{\ourmethod} is summarized in Algorithm \ref{alg:aidrag}, where Line 4$\sim$12 shows the process of $isRetrieve$ and Line 13$\sim$21 shows the process of $Select$.

\begin{algorithm}[H]
    \renewcommand{\algorithmicrequire}{\textbf{Input:}}
	\renewcommand{\algorithmicensure}{\textbf{Output:}}
    \renewcommand{\algorithmiccomment}[1]{\hfill \textcolor{blue}{$\triangleright$#1}}
	\caption{The complete process of \text{\ourmethod}} 
	\label{alg:aidrag} 
	\begin{algorithmic}[1]
		\REQUIRE the trained \textit{\ourmodule}, a code LM $\mathcal{\mathcal{M}_\theta}$, code to be completed $X$, the maximum number of iterations \textrm{MAX-ITER}, $T_{RAG}^i,T_{ACC}^i, i=1,\cdots,\textrm{MAX-ITER}$
        \STATE \textrm{Score} $\leftarrow$ []  \COMMENT{To save the predicted score $\hat{s}^i$ in each iteration $i$.}
        \STATE BestY $\leftarrow$ [] \COMMENT{To save the generation result $\mathbf{\hat{y}}^i$ in each iteration $i$.}
        \STATE $X^0$ $\leftarrow$ X  \COMMENT{Initialize $X^0$ with original $X$}
        
        \emph{Adaptive retrieval:}
        \FOR{$i$ := $0$ to \textrm{MAX-ITER}}
        \STATE Generate $\mathbf{\hat{y}^i}$ and $\hat{logits}^i$ using $\mathcal{M}_\theta$ for $X^i$.
        \STATE Extract the feature vector $z^i$ from $\hat{logits}^i$ as described in Section \ref{sec:aider}.
        \STATE Calculate the score: $\hat{s}^i\leftarrow\textit{\ourmodule}(z)$, and append $\hat{s}^i$ to \textrm{Score}
        \IF{not $isRetrieve(\hat{s}^i, T_{RAG}^i)$}
        \STATE \textbf{goto} \textit{Selective accept}
        \ENDIF
        \STATE Retrieve and update the prompt $X^{i+1}$ with newly retrieved code snippets and $X$
        \ENDFOR
        
        \textit{Selective accept:}
        \FOR{$i$ := $0$ to |\textrm{Score}|}
        \STATE Append $\mathbf{\hat{y}^i}$ to \textrm{BestY}
        \FOR{$j$ := $i-1$ to $0$}
        \IF{$Select(\textrm{Score}[j], \textrm{Score}[i], T_{ACC}^j)$}
        \STATE \textrm{BestY[i]} $\leftarrow$ \textrm{BestY[j]}
        \STATE \textbf{break} 
        \ENDIF
        \ENDFOR
        \ENDFOR
    \ENSURE \textrm{BestY}[-1] \COMMENT{Return the last value of \textrm{BestY}}
	\end{algorithmic} 
\end{algorithm}

\section{Experimental Setup}

\subsection{Evaluation Datasets}
To evaluate performance on various code complication tasks, we use the benchmark RepoEval~\cite{zhang2023repocoder}. It is a benchmark consisting of line, API, and function completion tasks from 14 repositories in Python. The line and API datasets contain 1,000 samples respectively, and the function dataset contains 373 samples. 
To investigate the generalization to other programming languages, we additionally construct a line completion benchmark named \text{\benchmark}. 
As Table~\ref{tab:bench_repos} shows,
we select 8 repositories from GitHub for Python, Java, JavaScript, and C, and 2 repositories are selected for each language. For each repository, we create 500 samples, thus 4,000 samples in total are included in \text{\benchmark}.

\begin{table}[H]
    \caption{The repositories selected to construct \text{\benchmark}. \# represents number. The repositories are selected following two criteria: 1) created after July 1, 2023, and 2) with more than 500 stars.}
    \label{tab:bench_repos}
    \centering
    \begin{tabular}{c|c|c|c}
    \toprule
        \textbf{Name} & \textbf{License} & \textbf{Created time} & \textbf{\# of files} \\
    \midrule
        \multicolumn{4}{c}{\cellcolor{black!20}Language: C}       \\
        raddebugger\footnotemark[1] & MIT License& 2024-01-10 & 140 \\
        valkey\footnotemark[2] & BSD-3 Clause& 2024-03-22 & 429  \\
        \multicolumn{4}{c}{\cellcolor{black!20}Language: Java}       \\
        conductor\footnotemark[3] & Apache V2.0 & 2023-12-08 & 675 \\
        bindiff\footnotemark[4] & Apache V2.0 & 2023-09-20 & 1,045 \\
        \multicolumn{4}{c}{\cellcolor{black!20}Language: JavaScript}       \\
        puter\footnotemark[5] & AGPL-3.0& 2024-03-03& 672\\
        AD\_Miner\footnotemark[6] & GPL-3.0& 2023-09-26& 15\\
        \multicolumn{4}{c}{\cellcolor{black!20}Language: Python}       \\
        SWIFT-AI\footnotemark[7] & Apache V2.0 & 2023-07-11 & 1,201\\
        IDM-VTON\footnotemark[8]& CC BY-NC-SA 4.0 & 2024-03-20 & 559 \\
        
    \bottomrule
    \end{tabular}
\end{table}
\footnotetext[1]{\url{https://github.com/EpicGamesExt/raddebugger}}
\footnotetext[2]{\url{https://github.com/valkey-io/valkey}}
\footnotetext[3]{\url{https://github.com/conductor-oss/conductor}}
\footnotetext[4]{\url{https://github.com/google/bindiff}}
\footnotetext[5]{\url{https://github.com/HeyPuter/puter}}
\footnotetext[6]{\url{https://github.com/Mazars-Tech/AD_Miner}}
\footnotetext[7]{\url{https://github.com/liwenxi/SWIFT-AI}}
\footnotetext[8]{\url{https://github.com/yisol/IDM-VTON}}
 

\subsection{Evaluation Metrics}
For line and API completion tasks, we evaluate the generated code with two metrics: Exact Match (EM) and Edit Similarity (ES).
For the function completion task, we evaluate the generated function with unit test pass rate (UT) and ES. 
Each sample is regarded as passed when all unit tests succeed (scored 1) otherwise failed (scored 0).

\subsection{Target RAG-based code completion system}
We follow the RAG framework proposed by RepoCoder~\cite{zhang2023repocoder}.

\textbf{Retriever.} Repocoder uses sliding windows to cut the code files into code snippets. We set the window sizes to 50 lines for function completion and 20 for others (\text{\benchmark} and line, API completion for RepoEval). The sliding stride is set to 10. The returned code snippets are appended to the prompt following \citet{zhang2023repocoder} until the number of tokens exceeds 4k.
   
\textbf{Generators.}
We use two popular open-sourced code LMs for our evaluation: CodeLlama-7B~\cite{rozière2024code} and DeepSeek-Coder-7B~\cite{guo2024deepseekcoder}.
We set the number of tokens of $X$ to 1.2k for zero-shot generation, and 4k for \RG as mentioned before.
The maximum number of generated tokens is set to 50 for line~/~API completion and 300 for function completion. 
We operate post-processing for the generated code.
For line and API completion, we truncate the prediction to the same number of lines as in ground truth. 
For function completion, we extract the function body by matching the number of left and right brackets.

\subsection{\text{\ourmethod}} \label{sec:setups}
\textbf{Training \textit{\ourmodule}.} To train our \textit{\ourmodule} (Section \ref{sec:aider}), we need to provide a collection of data pairs containing predictions and its corresponding logits generated by the code LM.  We first construct a dataset for code completion using the Stack~\cite{kocetkov2022stack} dataset. In detail, we sample 11k Python repositories which have more than 50 and less equal to 100 files from the Stack. For each repository, we select the files with at least 3 local imports and more than 20 non-empty lines of code as candidates. We sample $(X, y)$ pairs and employ the K-Means \cite{MacQueen1967SomeMF} to remove similar pairs following \citet{wu2024repoformer}. The number of lines of $y$ follows Poisson distribution ($P(X=k)={\frac{e^{-\lambda }\lambda ^{k}}{k!}}$), where $\lambda$ is set to 2. The number of lines of $X$ is set to 50. In the end, we construct a total of 250k pieces of data collection.
Then, we use CodeLlama-7B~\cite{rozière2024code} and DeepSeek-Coder-7B~\cite{guo2024deepseekcoder} to complete the collection in a zero-shot way. We truncate $X$ to 1.2k tokens, and the maximum number of generated tokens is set to 50. 
LightGBM is built via the Python library \texttt{lightgbm} with the default parameters, and trained on the constructed $\mathcal{D}$.

\textbf{$\textbf{CARD-RG}_i$}. $\textit{CARD-RG}_i$ is the result of adapting our \text{\ourmethod} to one-iteration of RAG ($\textit{CARD-RG}_1$) and more iterations of RAG ($\textit{CARD-RG}_{2\sim4}$). 
    To obtain $\textit{CARD-RG}_1$, zero-shot generation is the input to \textit{\ourmodule}, and \text{\ourmethod} helps adaptive retrieval and selectively choose the results between $RG_1$ and zero-shot results. 
    For $i>2$, $\textit{CARD-RG}_i$ is conducted as a continuation of $\textit{CARD-RG}_{i-1}$. In other words, the samples decided not to be retrieved in the early stage will not be judged in the later stage. However, $\textit{CARD-RG}_2$ is based on $RG_1$ rather than $\textit{CARD-RG}_1$, as $RG_1$ is the first iteration instead of the zero-shot stage for iterative RAG.

\textbf{Thresholds.} The thresholds $T_{RAG}$ and $T_{ACC}$ are different for each iteration. For all line-level completion tasks (line and API for RepoEval and all datasets for \text{\benchmark}), $T_{RAG}$ is set to $0.9, 0.8, 0.7,$ $ 0.6$ for $\textit{CARD-RG}_i$, $i=1,2,3,4$ respectively, and $T_{ACC}$ are set to $0.8, 0.9,$ $ 0.95$ and $0.99$. For function completion, $T_{RAG}$ are set to $[0.65, 0.45, 0.3, 0.25]$ and $T_{ACC}$ are set to $[0.9, 0.9, 0.95, 0.99]$.

\section{Results} \label{sec:results}
In this section, we show the experimental results on RepoEval and \text{\benchmark}, analyze why \text{\ourmethod} works, and illustrate the experimental findings.

\subsection{Is \text{\ourmethod} beneficial to RAG-based code completion task?}
\label{sec:main_result}

\subsubsection{Performance}

\begin{table*}[htbp]
\caption{Results on RepoEval. RG$i$ represents the $i$-th iteration of RAG, and \text{\ourmethod}$i$ represents the $i$-th iteration of \text{\ourmethod}. The accumulative Average Retrieval Times (aART) reported, and the values in the brackets are the reduced ratio of aART compared with invariable retrieval. ART equals to total times of retrieval divided by the number of samples.}
\label{tab:results_main}
\resizebox{\linewidth}{40mm}{
\begin{tabular}{c|c|c|c|c|c|c|c|c|c|c}
\toprule
\textbf{Dataset} & \textbf{Metric} & \textbf{Zero-shot} & \textbf{$\textit{RG}_1$} & \textbf{$\textit{CARD-RG}_1$} &\textbf{$\textit{RG}_2$} & \textbf{$\textit{CARD-RG}_2$} & \textbf{$\textit{RG}_3$} & \textbf{$\textit{CARD-RG}_3$} & \textbf{$\textit{RG}_4$} & \textbf{$\textit{CARD-RG}_4$} \\ 

\midrule
 \multicolumn{11}{c}{\cellcolor{black!20}CodeLlama-7B}                   \\
\multirow{3}{*}{\textbf{Line}} 
& EM     &   33.94\% & 52.31\%  &  52.56\% & 53.19\%  &   53.81\%  & 53.75\%   &  54.25\%& 54.06\%  & 54.25\%         \\
& ES     &     59.42\%   & 71.83\% &  72.26\% & 72.47\%  &  73.03\%   & 73.03\%    & 73.35\%    & 73.26\%  & 73.37\% \\
& aART & 0 & 1 &  0.79(-21.0\%) & 2 & 1.53(-23.5\%) & 3 & 1.92(-36.0\%) & 4 & 2.21(-44.8\%) \\
\midrule
\multirow{3}{*}{\textbf{API}} 
& EM     &  25.31\% & 40.50\% & 40.38\% & 41.44\%  &  42.56\% & 42.19\%  &  42.94\% & 41.94\%  & 43.19\% \\
& ES     &  54.82\% & 66.90\% & 67.01\% & 67.43\% & 68.23\% & 67.89\% & 68.56\% & 67.71\% & 68.58\% \\
& aART & 0 & 1 &  0.82(-19.0\%) & 2 & 1.60(-20.0\%) & 3 & 2.08(-30.7\%) & 4 & 2.45(-38.8\%) \\
\midrule
\multirow{3}{*}{\textbf{Function}}
& UT     &  28.42\%  & 34.32\% & 35.12\%  & 35.39\%  & 36.46\% & 36.46\% & 37.00\% & 35.39\% & 37.00\% \\
& ES     & 38.62\%  & 48.79\% & 48.82\%  & 50.23\% & 50.40\% & 50.62\% & 50.68\% & 50.49\% &  50.72\%        \\
& aART & 0 & 1 &  0.94(-6.0\%) & 2 & 1.75(-12.5\%) & 3 & 2.11(-29.7\%) & 4 & 2.21(-44.8\%)\\

 \multicolumn{11}{c}{\cellcolor{black!20}DeepSeek-Coder-7B}                                                      \\
\multirow{3}{*}{\textbf{Line}}  & EM     & 36.25\%  & 54.56\%  & 54.87\%  & 55.75\% &  56.63\%  & 56.56\%  & 57.06\%  & 56.13\% & 57.13\%  \\
& ES     &  60.98\%  & 73.23\%  & 73.59\% & 74.30\% & 74.97\% & 74.73\% & 75.21\% & 74.44\%  &  75.24\%  \\
& aART & 0 & 1 &  0.77(-23.0\%) & 2 & 1.51(-24.5\%) & 3 & 1.89(-37.0\%) & 4 & 2.16(-46.0\%) \\
\midrule
\multirow{3}{*}{\textbf{API}} & EM     &  26.38\% & 42.38\% & 42.44\%& 44.25\% & 45.06\% & 44.75\% & 45.25\% & 44.75\%  & 45.50\% \\
& ES     &  56.79\% & 68.72\% & 68.93\% & 70.02\% & 70.55\% & 70.69\% & 71.02\% & 70.33\%  & 71.22\% \\
& aART & 0 & 1 &  0.86(-14.0\%) & 2 & 1.59(-20.5\%) & 3 & 2.04(-32.0\%) & 4 & 2.40(-40.0\%) \\
\midrule

\multirow{3}{*}{\textbf{Function}} & UT     &  28.42\% & 36.46\% & 37.80\% & 35.93\% & 36.46\% & 37.27\% & 38.00\% & 37.27\% & 38.00\%         \\
& ES     & 39.04\% & 49.54\% & 49.54\% &  50.67\% & 51.06\% & 50.76\%  & 51.21\% & 50.92\% & 51.27\% \\
& aART & 0 & 1 &  0.93(-7.0\%) & 2 & 1.72(-14.0\%) & 3 & 2.08(-30.7\%) & 4 & 2.14(-46.5\%) \\
\bottomrule
\end{tabular}
}
\end{table*}

We show the evaluation results of CodeLlama-7B and DeepSeek-Coder-7B on RepoEval in Table \ref{tab:results_main}. 
From Table~\ref{tab:results_main}, we see that \text{\ourmethod} saves 6\% to 46.5\% times of RG, while improving ES on all setups. We list some interesting results as follows.
 
Regards to tasks, \text{\ourmethod} saves 21\% to 46\% times of RG for Line completion, 14\% to 40\% times of RG for API completion, and 6\% to 46.5\% times of RG for function completion respectively, while improving the ES on all setup. 

Regards to iterations, \text{\ourmethod} saves 6\%$\sim$23\% for the first iteration, 12.5\%$\sim$24.5\% for the second iteration, 29.7\%$\sim$37.0\% for the third iteration, and 38.8\%$\sim$46.5\% for the fourth iteration. The $\textit{CARD-RG}_4$ performs the best for all settings in terms of all metrics. The $\textit{CARD-RG}_2$ performs better than all RG1 to RG4 for line and API tasks, with only $1.50\sim 1.60$ times of $RG$. As expected, $\textit{CARD}$ maintains a non-declining trend as more iterations are applied due to our selective acceptance policy, while the original iterative solution may decline on ES ($RG_3$ to $RG_4$) due to \textit{degenerated cases} problem. Notably, the $\textit{CARD-RG}_4$ introduces only less than 0.1 additional times of RG while still improving ES for function completion.

Considering the scenario with similar time costs, the $\textit{CARD-RG}_3$ applies similar times of RG compared to $RG_2$, but achieves more than 1\% improvement on EM for line and API completion, and up to 2.1\% improvement on UT for function completion. 

In total, through the whole process of \text{\ourmethod}, the ES is improved by 12.23\%$\sim$14.43\% and 1.54\%$\sim$2.50\% compared to zero-shot and $\RG_1$ generation.

\subsubsection{Latency}

\begin{table}[ht]
    \centering
    \caption{The reduced latency (RL) for each repository for \text{\benchmark} and the average RL.}
    \label{tab:latency1}
    \begin{tabular}{c|c|c|c|c|c}
    \toprule
        \multirow{2}{*}{\textbf{Repository}} &\multirow{2}{*}{$\mathbf{T_r}$} & \multicolumn{2}{c|}{\textbf{$\textit{CARD-RG}_1$}} & \multicolumn{2}{c}{\textbf{$\textit{CARD-RG}_2$}}\\
        \cline{3-6}
        & & ART & RL & ART & RL\\
        \midrule
        raddebugger & 728 & 74.6\% & 13.1\% & 43.2\% & 56.8\%\\
        valkey & 1664 & 73.4\% & 10.1\% & 58.2\% & 41.8\%\\
        conductor & 623 & 59.6\% & 16.8\% & 22.8\% & 77.2\%\\
        bindiff & 613 & 62.8\% & 14.3\% & 30.4\% & 69.6\%\\
        puter & 845 & 57.8\% & 23.1\% & 30.6\% & 69.4\%\\
        AD\_Miner & 843 & 64.4\% & 19.5\% & 48.2\% & 51.8\%\\
        SWIFT-AI & 837 & 73.0\% & 14.8\% & 44.0\% & 56.0\%\\
        IDM-VTON & 774 & 71.0\% & 16.5\% & 26.0\% & 74.0\% \\
        \midrule
        \textbf{Average} & 866 & 67.1\% & 16.0\% & 37.9\% & 62.1\% \\
    \bottomrule
    \end{tabular}
\end{table}

We analyze the concrete latency reduced by our adaptive retrieval in \text{\ourmethod} in the realistic scenario. 
For each retrieval-generation-decision process, we consider three latency terms: 
(1) $T_d$, the time required for decision-making by \text{\ourmethod}, 
(2) $T_r$, the retrieval latency, 
and (3) $T_{gi}$, the latency for the $i$-th generation (i starts from 0).

The total latency $T_i$ for a sample for $\textit{CARD-RG}_i$ is formulated as:
\begin{equation*}
    T_i = \begin{cases}
        T_d&, \text{without\ retrieval}\\
        T_d+T_r+T_{gi}&, \text{with\ retrieval}
    \end{cases}
\end{equation*}

For single RAG introduced in Section \ref{sec:application}, the generation and retrieval can be conducted in parallel. Thus, $T_r$ is reduced for samples with retrieval, and $\max(T_{g0}, T_r)$ is added for all samples.

We benchmark the average latency of $T_{r}$, $T_{g0}$ and $T_{gi},i>0$ for CodeLlama-7B with the vLLM library~\cite{kwon2023efficient} on a single Nvidia A100 GPU (40G). $T_{d}$ is negligible compared with other items, which is about 1 millisecond.
$T_{g0}$ and $T_{gi},i>0$ are 755 and 1,025 milliseconds respectively. $T_r$ and the calculated reduced latency for $\textit{CARD-RG}_1$ and $\textit{CARD-RG}_2$ following the equations above are listed in Table \ref{tab:latency1}. The results show that \text{\ourmethod} could reduce 16\% latency for $RG_1$, and 62\%, 72\%, and 83\% for $RG_{2\sim4}$ in the realistic scenario. 

From the above results, \text{\ourmethod} is empirically proved to effectively alleviate the \textit{degenerated cases}, and reduce the redundant RG processes. By measuring the latency in the realistic scenario, \text{\ourmethod} shows 16\%$\sim$83\% time-saving for each iteration.

\subsection{How about the generalizability of \text{\ourmethod}?}
\label{sec:generalizable}

In this section, we conduct experiments to evaluate \text{\ourmethod}'s generalization ability in 4 different dimensions.

\subsubsection{Different Retrievers}
\begin{figure}
    \centering
    \includegraphics[width=\linewidth]{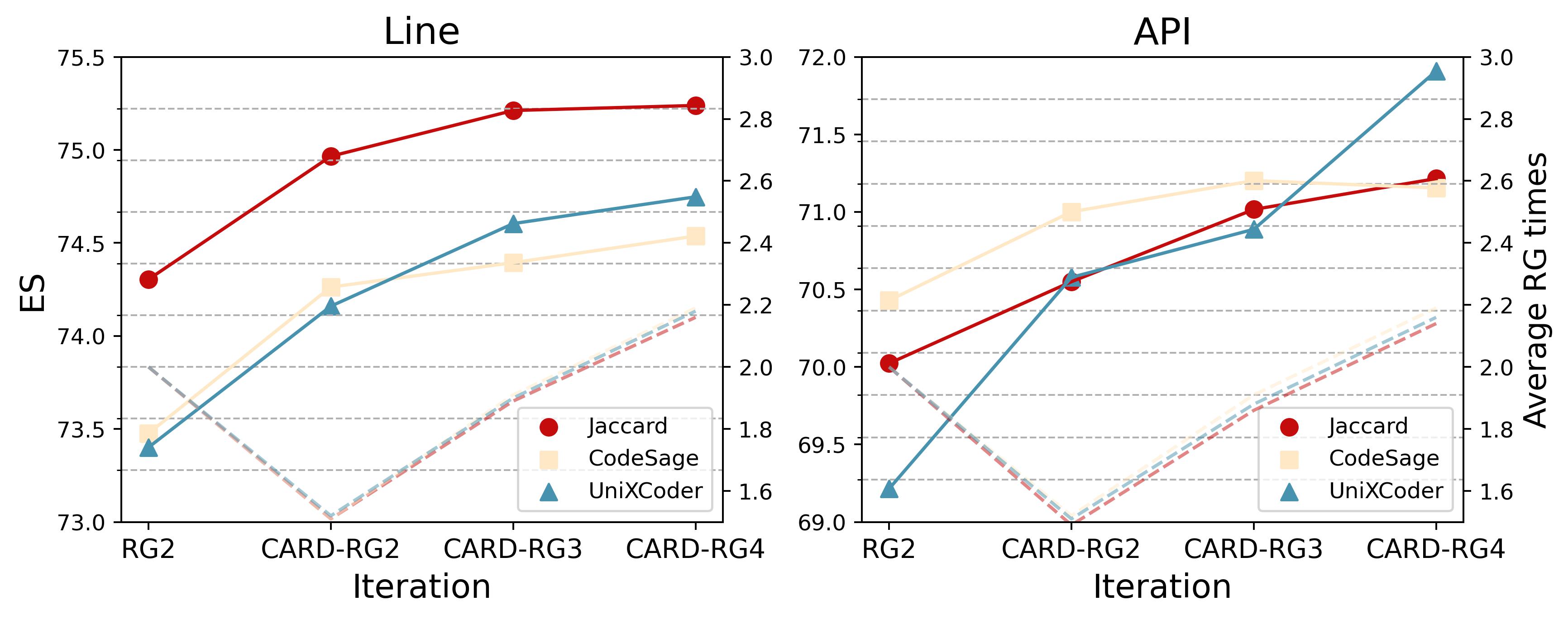}
    \caption{Performance under different retrievers when the code LM is DeepSeek-Coder-7B.}
    \label{fig:retrievers}
\end{figure}

Intuitively, our \textit{\ourmodule} only accepts the output of LMs, thus would not be impacted by the retrieval setups (e.g. different retrievers). To experimentally prove it, apart from the Jaccard similarity as the retriever, we introduce two dense retrievers, UniXCoder~\cite{guo2022unixcoder} and CodeSage~\cite{zhang2024codesage} to evaluate \text{\ourmethod} with different retrieval setups. 
We replace Jaccard with these two retrievers and utilize DeepSeek-Coder-7B as the code LM. 
The results are summarized in Figure \ref{fig:retrievers}. The solid and dashed lines represent the ES value and average RG times respectively.

Though different retrievers show different ES values, the non-declined trend of ES remains observed for all retrievers except $\textit{CARD-RG}_4$ for CodeSage, and $\textit{CARD-RG}_2$ shows steady improvement compared with $RG_2$. The average $RG$ times also show consistency among retrievers, indicating that our \text{\ourmethod} is retriever-insensitive and well applicable for different retrieval setups.

\subsubsection{Different generators}

\begin{table}
\caption{CARD on different code LMs. Only ES is reported in the table. CL, DS, SC represents CodeLlama, DeepSeekCoder, and StarCoder2 respectively. The values in the bracket are the aARTs.}
\label{tab:results_codelm}
\centering
\begin{tabular}{c|c|c|c|c|c}
\toprule
\textbf{Series} & \textbf{Zeroshot} & \textbf{$\textit{RG}_1$} & \textbf{$\textit{CARD-RG}_1$} & \textbf{$\textit{RG}_2$} & \textbf{$\textit{CARD-RG}_2$} \\ 

\midrule
\textbf{CL-13B} &  60.30\% & 72.13\%  & 72.45\%(0.79) &  73.03\% & 73.65\%(1.51)  \\
\midrule
\textbf{DS-1.3B} & 56.85\% &  70.71\% & 70.71\%(0.84)  & 71.53\% & 72.00\%(1.56)    \\
\midrule
\textbf{SC-3B} & 58.79\% & 72.25\%  & 72.27\%(0.97) & 72.47\% & 72.89\%(1.87)      \\
\midrule
\textbf{SC-7B} & 59.86\% & 73.30\%   & 73.29\%(0.97)  & 73.67\% & 74.14\% (1.87)          \\
\midrule
\textbf{SC-15B} & 61.31\% & 74.24\% & 74.24\%(0.96)  & 74.51\%& 74.95\% (1.86)     \\
\bottomrule
\end{tabular}
\end{table}

\begin{figure}
    \centering
    \includegraphics[width=\linewidth]{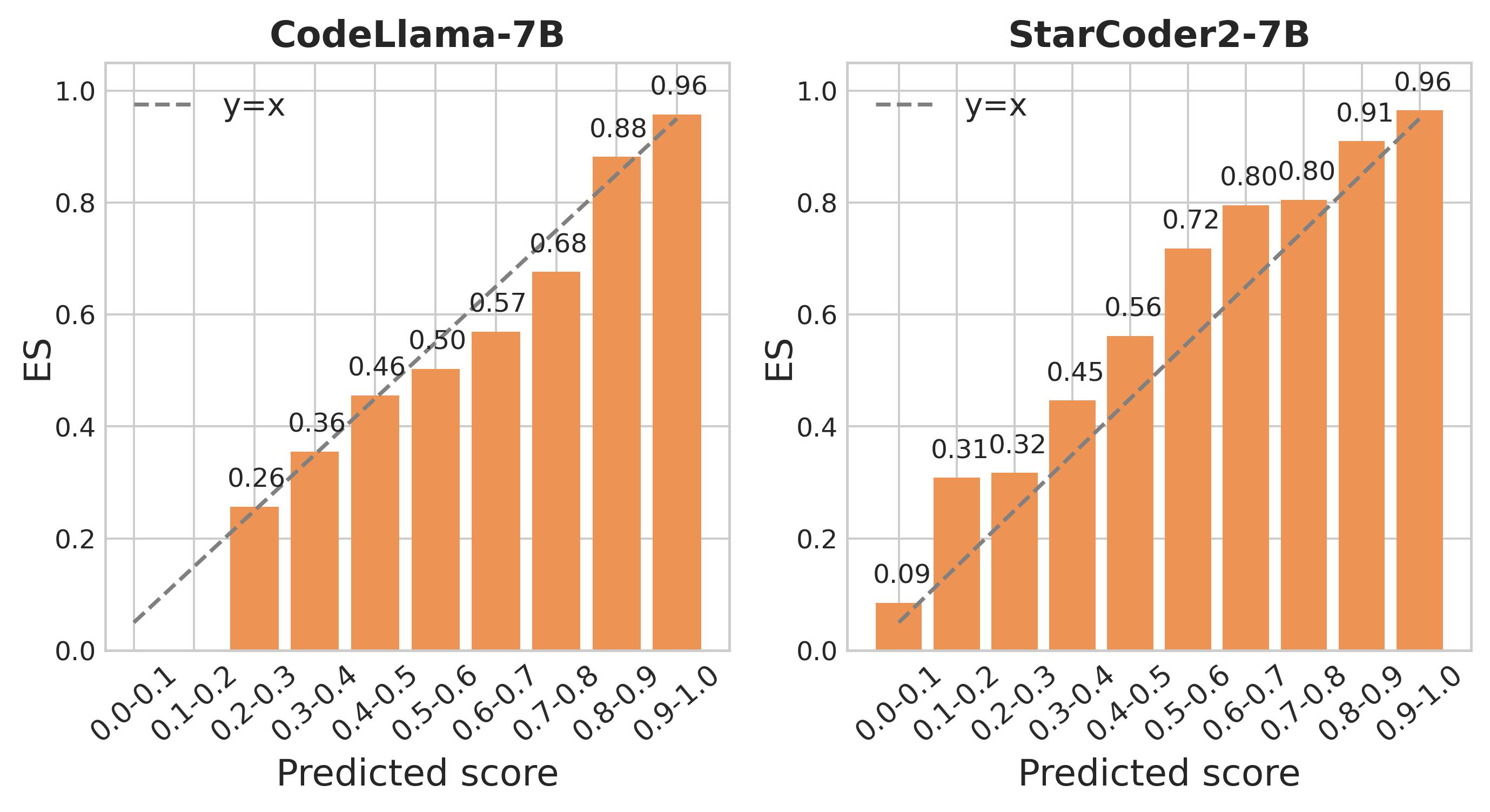}
    \caption{The ground truth ES versus the predicted ES. Each bar represents the average ES of the predicted values within the range on the x-axis.}
    \label{fig:codelm_relation}
\end{figure}

To evaluate our trained \textit{\ourmodule} on different code LMs, we select 5 different LMs apart from CodeLlama7B and DeepSeek-Coder-7B, listed as follows:

\textbf{Intra-family}: Code LMs that are in the same family as CodeLlama7B and DeepSeek-Coder-7B, including CodeLlama-13B and DeepSeek-Coder-1.3B.

\textbf{Inter-family}: A new family of Code LMs called StarCoder2~\cite{lozhkov2024starcoder}, including StarCoder2-Base-3B, StarCoder2-Base-7B and StarCoder2-15B.

The results on the line dataset of RepoEval are listed in Table~\ref{tab:results_codelm}. From Table~\ref{tab:results_codelm}, \textit{intra-family} code LMs also exhibit performance improvement and the average RG times are similar to results in \ref{tab:results_main}. 

However, a large shift occurs when the code LM belongs to the family of StarCoder2. Although the ES values of $\textit{CARD-RG}_2$ surpass that of $RG_2$, $\textit{CARD-RG}_1$ shows no obvious improvement. In addition, the retrieval ratio significantly increases to more than 0.95\% for $\textit{CARD-RG}_1$ and 1.85\% for $\textit{CARD-RG}_2$.
We show the relationship between ES and predicted score $\hat{s}$ in Figure \ref{fig:codelm_relation}, where the x-axis is the range of predicted score $\hat{s}$ and the y-axis is the average ES value of samples falling in each range of $\hat{s}$. We can see for CodeLlama-7B, $\hat{s}$ is closer to ES in statistical meaning (the top of the bar is closer to the dashed line $y=x$), while the right figure shows more under-confident $\hat{s}$. 
The results demonstrate that our method is also applicable when faced with an unseen code LM, but it is advised to refit \textit{\ourmodule} on the new LM.

\begin{table*}[ht]
\caption{Results on \text{\benchmark}.}
\label{tab:results_multilanguage}
\resizebox{\linewidth}{50mm}{
\begin{tabular}{c|c|c|c|c|c|c|c|c|c|c}
\toprule
\textbf{Dataset} & \textbf{Metric} & \textbf{Zero-shot} & \textbf{$\textit{RG}_1$} & \textbf{$\textit{CARD-RG}_1$} &\textbf{$\textit{RG}_2$} & \textbf{$\textit{CARD-RG}_2$} & \textbf{$\textit{RG}_3$} & \textbf{$\textit{CARD-RG}_3$} & \textbf{$\textit{RG}_4$} & \textbf{$\textit{CARD-RG}_4$} \\ 

\midrule
 \multicolumn{11}{c}{\cellcolor{black!20}CodeLlama-7B}      \\
\multirow{3}{*}{\textbf{Python}} & EM     & 47.00\% & 69.20\% & 69.50\% & 69.20\%  & 70.10\% & 69.70\% & 70.20\% & 69.80\% & 70.20\%    \\
& ES     & 68.21\% & 81.27\% & 81.71\% & 81.50\% & 81.74\% & 81.82\%  & 82.05\% & 81.82\% & 82.07\%    \\
& aART    & 0 & 1 & 0.72(-28.0\%) & 2 & 1.35(-32.5\%) & 3 & 1.58(-47.3\%) & 4& 1.75(-56.3\%)\\
\midrule
\multirow{3}{*}{\textbf{C}} 
& EM     & 44.10\% & 55.40\% & 55.40\% & 55.00\% & 56.00\% & 55.10\% & 56.00\% & 55.50\% & 56.30\%   \\
& ES     & 67.32\% & 74.48\% & 74.63\% & 74.44\% & 75.02\% & 74.83\% & 75.07\% & 74.75\% & 75.12\%   \\
& aART    & 0 & 1 & 0.74(-26.0\%) & 2 & 1.51(-24.5\%) & 3 & 1.87(-37.7\%) & 4 & 2.11(-47.3\%)\\
\midrule
\multirow{3}{*}{\textbf{Java}} 
& EM     & 60.10\% & 76.10\% & 76.30\%  & 77.30\% & 78.60\% & 77.20\% &  78.60\% & 77.10\% & 78.60\%    \\
& ES     & 77.85\% & 86.28\% & 86.45\%  & 87.04\% & 87.82\% & 87.00\% & 87.82\% & 86.90\% & 87.82\%  \\
& aART    & 0 & 1 & 0.61(-39.0\%) & 2 & 1.27(-36.5\%) & 3 & 1.43(-52.3\%) & 4 & 1.52(-62.0\%) \\
\midrule
\multirow{3}{*}{\textbf{JavaScript}}
& EM     & 52.20\% & 61.10\% & 61.20\%  & 61.70\% & 62.30\% & 62.10\% & 62.60\% & 61.80\% & 62.70\% \\
& ES     & 71.75\% & 77.51\% & 77.67\% & 78.12\% & 78.39\% & 78.60\% & 78.69\% & 78.26\% & 78.71\% \\
& aART    & 0 & 1 & 0.60(-40.0\%) & 2 & 1.43(-28.5\%) & 3 & 1.76(-41.3\%) & 4 & 2.00(-50.0\%) \\
\midrule

\multicolumn{11}{c}{\cellcolor{black!20}DeepSeek-Coder-7B}                                             \\
\multirow{3}{*}{\textbf{Python}}  & EM     & 46.50\% & 70.30\%  & 70.30\%  & 71.10\% & 71.60\% & 71.20\% & 71.80\% & 71.20\% & 71.80\%         \\
& ES     & 68.22\% & 82.05\%  & 82.32\% & 82.84\% & 83.14\% & 82.89\% & 83.16\% & 83.13\% & 83.21\%   \\
& aART    & 0 & 1 & 0.75(-25.0\%) & 2 & 1.32(-34.0\%) & 3 & 1.55(-48.3\%) & 4& 1.68(-58.0\%)\\
\midrule
\multirow{3}{*}{\textbf{C}}                                            
& EM     & 40.50\% & 52.60\% & 53.60\% & 53.50\% & 54.20\% & 53.60\% & 54.40\% & 53.80\% & 54.50\% \\
& ES     & 64.71\% & 72.84\% & 73.37\% & 72.67\% & 73.55\% & 72.92\% & 73.87\% & 72.74\% & 73.95\% \\
& aART    & 0 & 1 & 0.75(-25.0\%) & 2 & 1.53(-23.5\%) & 3 & 1.95(-35.0\%) & 4 & 2.24(-44.0\%)\\
\midrule

\multirow{3}{*}{\textbf{Java}} 
& EM     & 59.40\% & 76.30\% & 76.40\% & 77.10\% & 78.50\% & 77.10\% & 78.70\% & 77.20\% & 78.70\%    \\
& ES     & 76.72\% & 86.54\% & 86.58\% & 87.02\% & 87.91\% & 86.97\% & 87.94\% & 87.04\% & 87.95\%     \\
& aART    & 0 & 1 & 0.62(-38.0\%) & 2 & 1.28(-36.0\%) & 3 & 1.43(-52.3\%) & 4 & 1.52(-62.0\%) \\
\midrule

\multirow{3}{*}{\textbf{JavaScript}}   
& EM     & 37.70\% & 53.00\% & 53.10\% & 53.40\% & 54.20\% & 53.50\% & 54.50\% & 53.60\% & 54.70\% \\
& ES     & 62.89\% & 72.66\% & 72.85\% & 73.15\% & 73.52\% & 73.28\% & 73.64\% & 73.35\% & 73.83\% \\
& aART    & 0 & 1 & 0.78(-22.0\%) & 2 & 1.56(-22.0\%) & 3 & 1.98(-34.0\%) & 4 & 2.29(-42.8\%) \\
\bottomrule
\end{tabular}
}
\end{table*}
\subsubsection{Multi languages}
From Table \ref{tab:results_multilanguage}, we see that \textit{\ourmodule} generalizes well on three other languages, though trained only with Python. 
The main findings in Table~\ref{tab:results_main} still hold: $\textit{CARD-RG}$ improves the performance and speed of RG and alleviates the problem of \textit{degenerated cases}. For all languages, $\textit{CARD-RG}_2$ averagely retrieves less than 1.5 times, achieving an EM even 1.5\% higher than $RG_4$ which retrieves 4 times. For $\textit{CARD-RG}_4$, the reduced ART ratio is up to 62\% for Java, implying that the ART for $\textit{CARD-RG}_4$ is even less than $RG_2$, while improving 1\% EM compared with $RG_2$. For $\textit{CARD-RG}_1$, 22\%$\sim$40\% of retrieval is rejected. Especially for JavaScript, up to 40\% of samples are decided not to RAG for CodeLlama-7B. 

Comparing different languages, we see that C and JavaScript have relatively high ART with \text{\ourmethod}. Generally, the higher performance $RG_4$ achieves, the more retrieval is rejected by \text{\ourmethod} with the same $T_{RAG}$. This phenomenon can be explained through the definition of $isRetrieve$: lower ES needs lower $T_{RAG}$ to maintain the same reduced retrieval ratio. 

Notably, the thresholds $T_{RAG}$ and $T_{ACC}$ are consistent with the setup for RepoEval, showing the robustness of our thresholds. The performance on \text{\benchmark} can be boosted with more fine-grained tuning on the specific programming language.

\subsubsection{Different \textit{\ourmodule}}

\begin{figure}[ht]
    \centering
    \subfigure[ES of generated code in each iteration for three machine-learning models.]{
    \label{subfig:ml_model1}
    \includegraphics[width=0.8\linewidth]{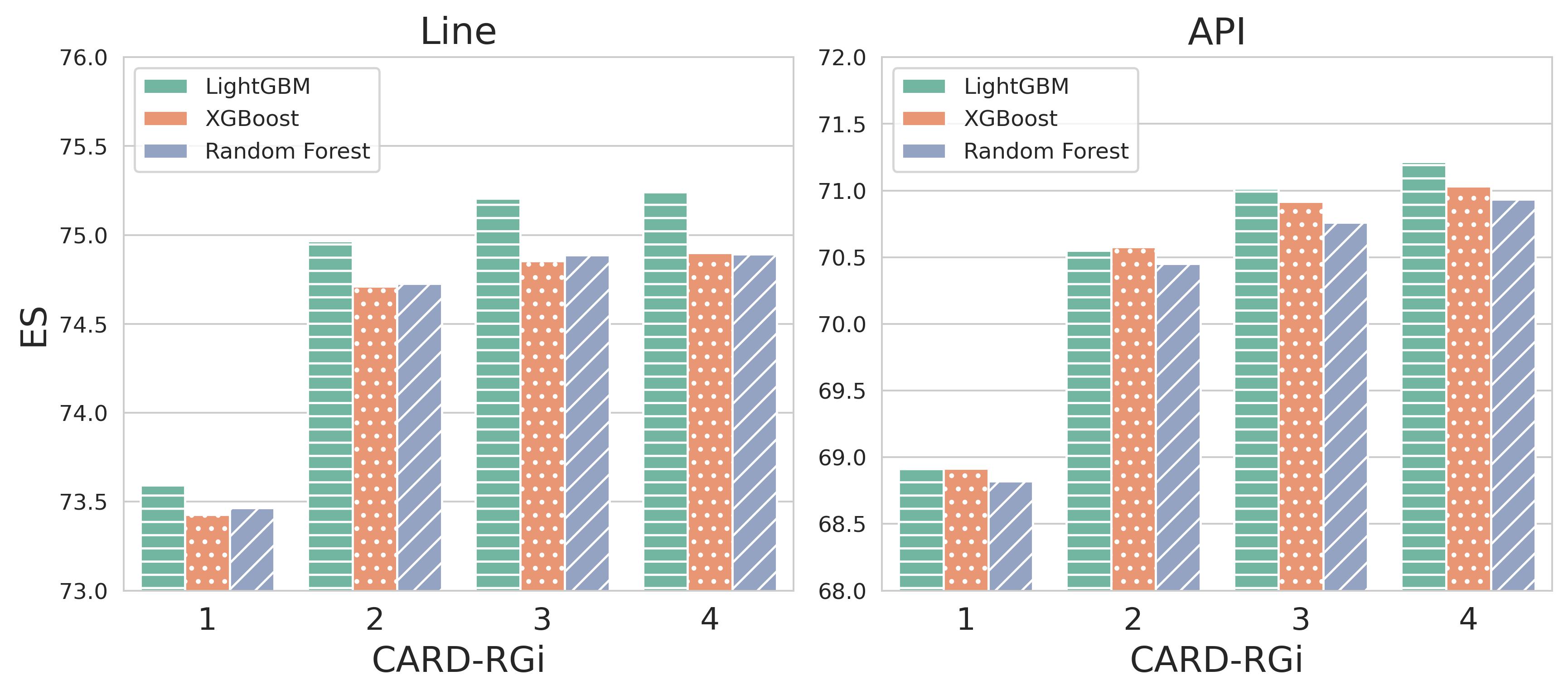}
    }
    \subfigure[Mean squared error of $\hat{s}$ and $ES$ in each iteration]{
    \label{subfig:ml_model2}
    \includegraphics[width=0.8\linewidth]{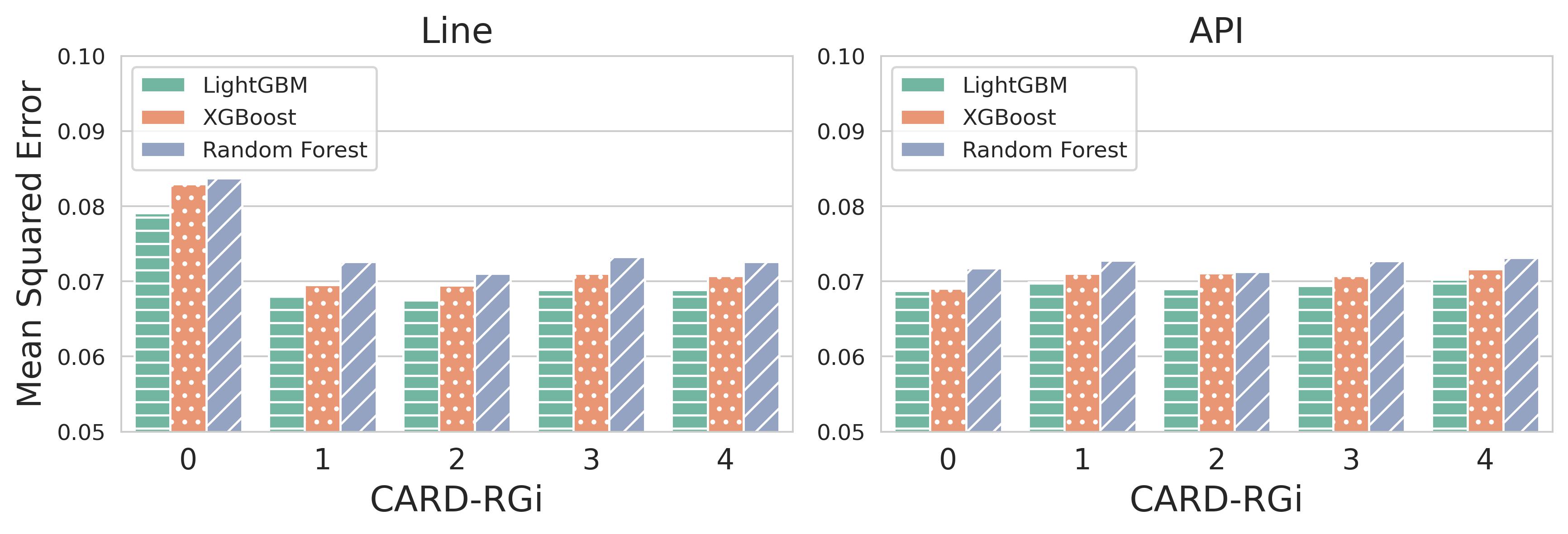}
    }
    \caption{Results on line dataset, where \textit{\ourmodule} is based on LightGBM, XGBoost, and Random Forest respectively, and the code LM is DeepSeek-Coder-7B.}
    \label{fig:ml_models}
\end{figure}

Besides LightGBM, XGBoost~\cite{Chen_2016} and Random Forest~\cite{breiman2001random} are also two widely used decision tree-based machine learning models. 
We train these models following the procedure described in Section \ref{sec:aider}, and integrate them into our \text{\ourmethod} to evaluate them. 
Those two models are implemented via Python library \textsc{XGBOOST} and \textsc{SCIKIT-LEARN} with the default parameters. 
The thresholds $T_{RAG}$ and $T_{ACC}$ follow the setups in Section \ref{sec:setups}. 

Figure~\ref{fig:ml_models} shows the end-to-end code completion performance (a) and the regression performance (b) on each iteration. 
Specifically, we calculate the mean squared error (MSE, lower MSE stands for better regression performance) of predicted $\hat{s}$ and the ground truth $s$ (which is also the ES value). 

As shown in Figure~\ref{subfig:ml_model1}, LightGBM outperforms the other two models slightly on both datasets. In addition, a non-declined trend is observed in terms of all three models, indicating that \text{\ourmethod} can be easily adapted to other machine learning models.
Comparing Figure \ref{subfig:ml_model1} and Figure \ref{subfig:ml_model2}, we see that the model achieving the lowest MSE (LightGBM) also achieves the best end-to-end performance, indicating that the performance of regression would significantly impact the end-to-end performance.
Based on it, we highlight the importance of \textit{\ourmodule} and point to a direction for obtaining better performance of \text{\ourmethod}.

In summary, this section shows the generalizability of \text{\ourmethod} from four aspects. The results show that our method can be combined with different RAG systems or different machine-learning models as \textit{\ourmodule}. When facing untrained scenarios (different programming languages and code LMs), \text{\ourmethod} still works in alleviating the problem of \textit{degenerated cases} and reducing the latency.

\subsection{Ablation study}
\label{sec:ablation}
We study some alternative setups in our \text{\ourmethod} from perspectives of feature engineering and design of \text{\ourmethod}:

\begin{table}[H]
\centering
\caption{Ablation study on RepoEval line and API dataset. Only ES is reported in the table, and the values in brackets are the difference with complete \text{\ourmethod}. The code LM adopted is DeepSeek-Coder-7B. }
\label{tab:results_ablation}
\begin{tabular}{c|c|c|c|c}
\toprule
\textbf{Setup} & \textbf{Dataset} & \textbf{$\textit{CARD-RG}_1$} & \textbf{$\textit{CARD-RG}_2$} & \textbf{$\textit{CARD-RG}_3$}\\ 

\midrule
\multirow{2}{*}{\textbf{FE1}} & Line     &  73.62(+0.03)  &   74.72(-0.25)  & 75.03(-0.18) \\      
& API     & 68.91(-0.02) & 70.57(+0.02)  & 70.95(-0.07)         \\
\midrule
\multirow{2}{*}{\textbf{FE2}} & Line     &   73.50(-0.09) & 74.97(-0.00) &   75.32(+0.11)    \\
& API     &  68.58(-0.35) & 70.56(+0.01)    &  70.99(-0.03)   \\
\midrule
\multirow{2}{*}{\textbf{D1}} & Line  & 73.37 (-0.22)  & 74.40 (-0.57) & 74.91 (-0.30) \\
& API     & 68.54 (-0.39) &  70.23 (-0.32)  & 70.90 (-0.12)               \\
\midrule
\multirow{2}{*}{\textbf{D2}} & Line     &   73.51 (-0.08)  &  75.04 (+0.07) &   75.34 (+0.13)    \\
& API     & 69.05 (+0.12) & 70.59 (+0.04) & 71.19 (+0.17) \\

\bottomrule
\end{tabular}
\end{table}

\begin{table}[H]
    \caption{The MSE loss with different ways of feature engineering. The lowest MSE for CodeLlama-7B and DeepSeek-Coder-7B are in \textbf{bold} and \underline{underlined} respectively. Reported in MSE $\times 10^3$.}
    \label{tab:mse_fe}
    \centering
    \begin{tabular}{c|c|c|c}
    \toprule
        \textbf{Setup} & \textbf{Code LM} & \textbf{MSE on Line} & \textbf{MSE on API} \\
    \midrule 
        \multirow{2}{*}{\textit{\ourmodule}} & \multirow{1}{*}{CodeLlama-7B} & \textbf{70.77} & \textbf{65.09}\\
        & \multirow{1}{*}{DeepSeek-Coder-7B} & \underline{70.87} & 62.40 \\
    \midrule
        \multirow{2}{*}{FE1} & \multirow{1}{*}{CodeLlama-7B} & 70.93 & 65.23\\
        & \multirow{1}{*}{DeepSeek-Coder-7B} & 70.98 & \underline{62.17}\\
    \midrule
        \multirow{2}{*}{FE2} & \multirow{1}{*}{CodeLlama-7B} & 72.53 & 65.32\\
        & \multirow{1}{*}{DeepSeek-Coder-7B} & 73.19 & 63.07\\
    \bottomrule
    \end{tabular}
\end{table}

\subsubsection{Feature Engineering}
We define \textbf{FE1} as only using probability-based features, and \textbf{FE2} as only using entropy-based features.

From Table \ref{tab:results_ablation}, we see that FE1 and FE2 perform similarly compared with our adopted feature engineering. However, the end-to-end ES is dependent on $T_{RAG}$ and $T_{ACC}$. We additionally calculate the average MSE of 5-iteration generations (zero-shot, $RG_{1\sim4}$) for each code LM on the line and API datasets listed in Table \ref{tab:mse_fe}. The results show that combining probability and entropy-based features is slightly superior to only using probability-based features and the entropy-based features are relatively less informative for regressing ES. For better feature extraction, our \textit{\ourmodule} can be combined with more semantic features (e.g. hidden states of LMs can be involved \cite{azaria2023internal, su2024unsupervised}), and even be trained as a deep neural network (such as long short-term memory \cite{6795963}) in an end-to-end way to handle the sequential input better.

\begin{figure}
    \centering
    \includegraphics[width=0.8\linewidth]{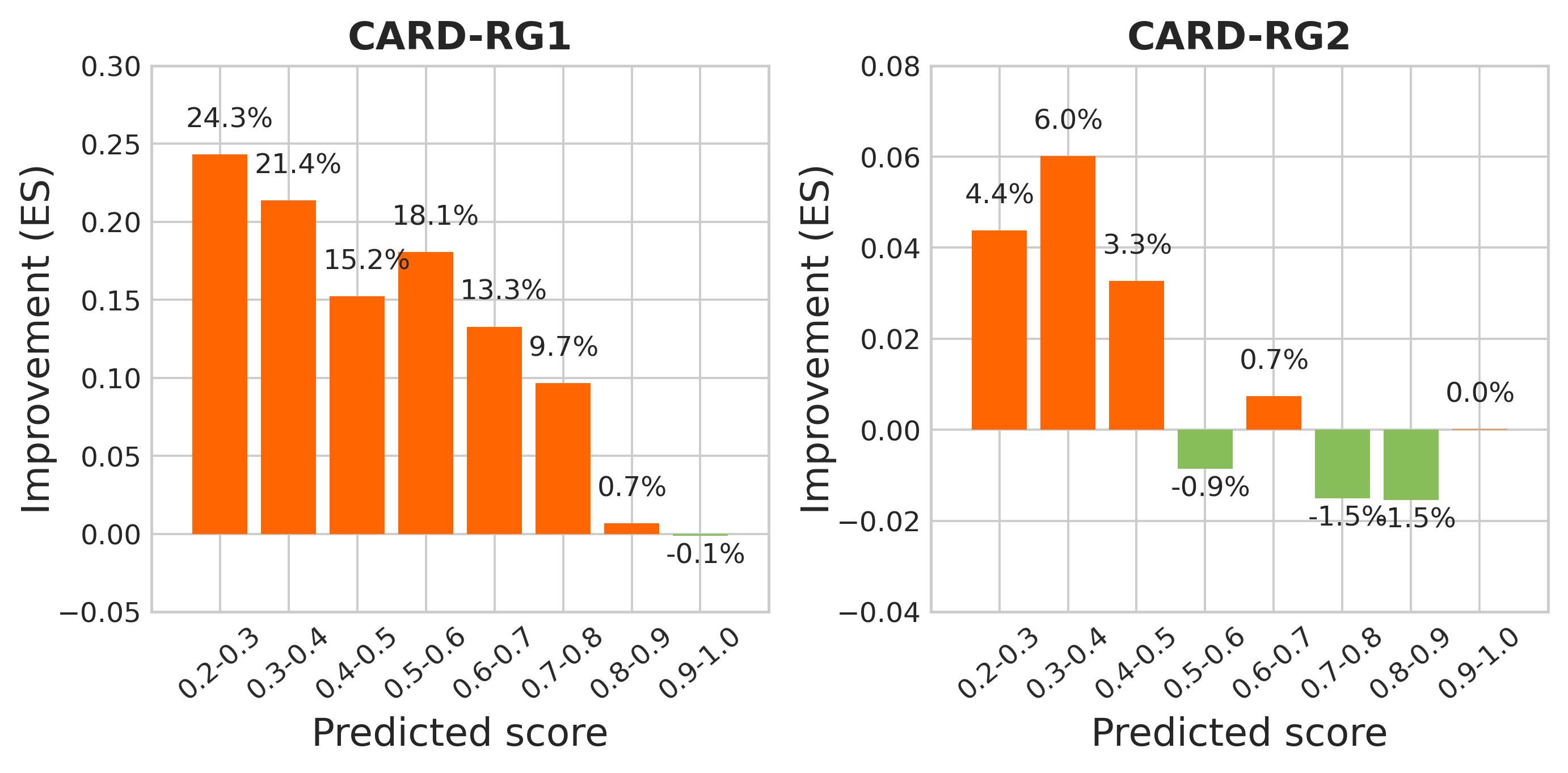}
    \caption{The improvement of ES under different ranges of predicted score $\hat{s}$ for $\textit{CARD-RG}_1$ and $\textit{CARD-RG}_2$. The improvement decreases dramatically as $\hat{s}$ increases.}
    \label{fig:esdiff_adarag}
\end{figure}

\subsubsection{Design of CARD}
\text{\ourmethod} can be divided into two separate functions: \textbf{D1} (only with \textit{Adaptive Retrieval} deciding whether to continue or conduct RAG) and \textbf{D2} (only with \textit{Selective Accept} selecting best result among iterations).

From Table \ref{tab:results_ablation}, we see that the performance of setup \textbf{D1} entirely decreases compared with complete \text{\ourmethod}, though it is still better than invariable RG. The performance of setup \textbf{D2} outperforms \text{\ourmethod} except $\textit{CARD-RG}_1$ on the API dataset, though in an invariable and time-consuming way of RAG. The results strengthen that the goal of adaptive retrieval and selective RAG are \textbf{speeding up} and \textbf{correcting the results} respectively. Intuitively, adaptive retrieval based on $\hat{s}$ relies on the correlation between ES and improvement of RAG, providing a solution to reduce redundant RAG and alleviate the \textit{degenerated cases} from a macro and coarse-grained perspective. From Figure \ref{fig:esdiff_adarag}, we see that the improvement shows a similar correlation to Figure \ref{fig:esdiff}, indicating that adaptive retrieval based on the predicted score of \textit{\ourmodule} is reasonable. Selective acceptance complements the adaptive retrieval through a fine-grained sample-level way to correct the \textit{degenerated cases}, regardless of the problem of latency. Combining these two mechanisms helps reduce unnecessary RAG and choose the best result from the existing generations.

\subsection{Hyper-parameters}
\label{sec:hyperp}
\begin{figure}
    \centering
    \includegraphics[width=0.8\linewidth]{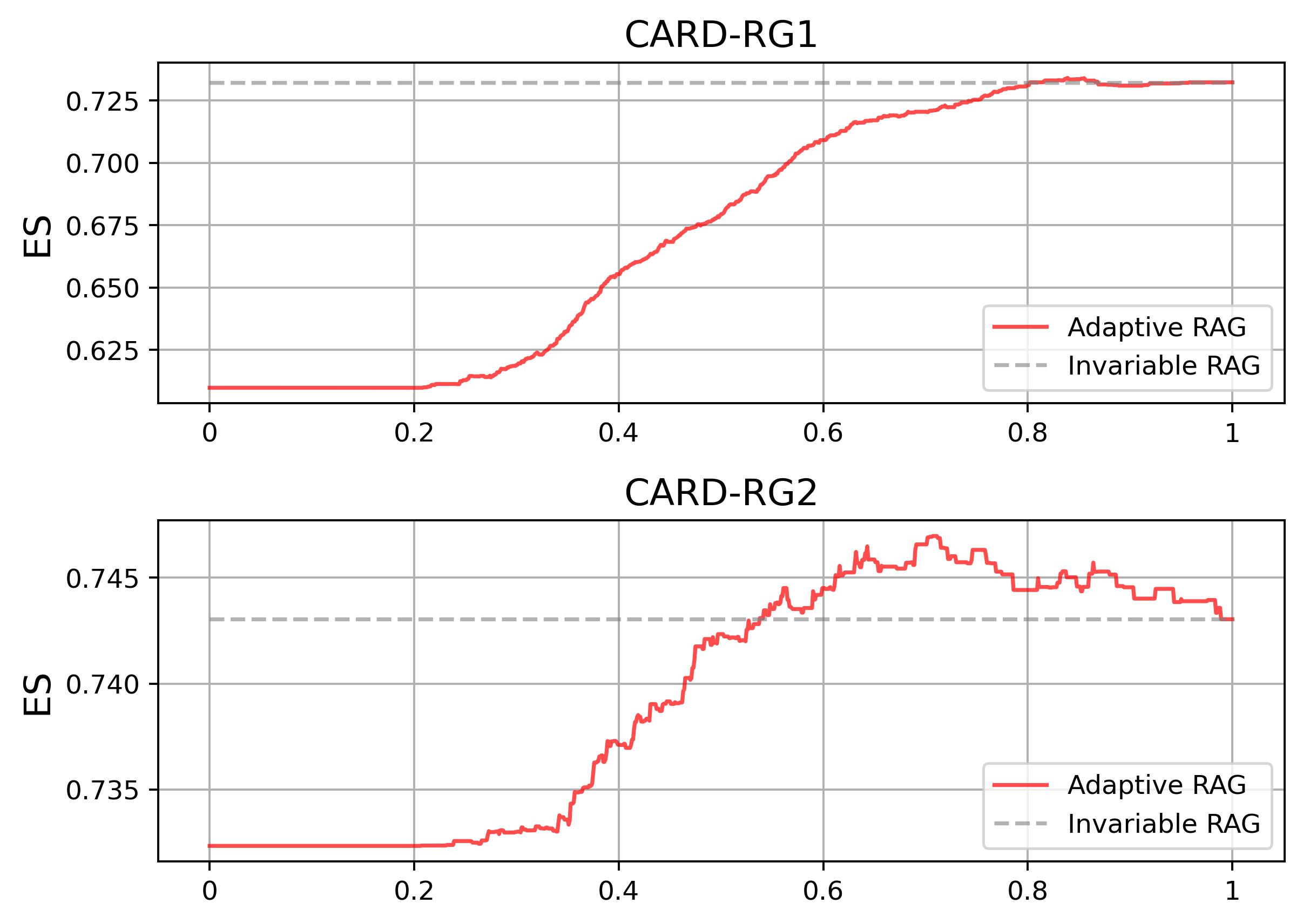}
    \caption{The change of ES as the $T_{RAG}$ increases. The dashed line represents the ES with invariable RG.}
    \label{fig:trag}
\end{figure}
\begin{figure}
    \centering
    \includegraphics[width=0.8\linewidth]{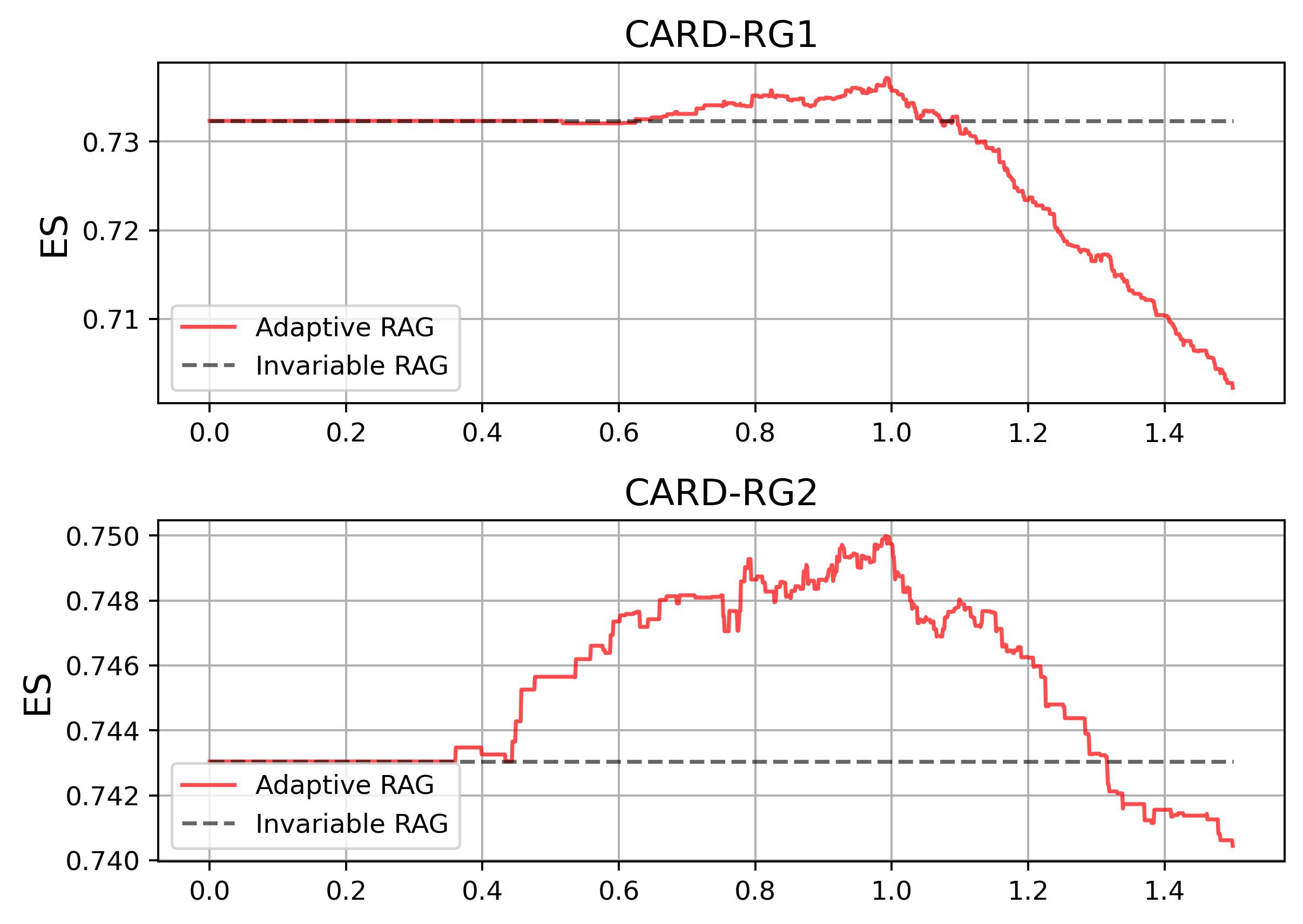}
    \caption{The change of ES as the $T_{ACC}$ increases. The dashed line represents the ES with invariable RG.}
    \label{fig:tacc}
\end{figure}

For the hyper-parameters involved in our method (such as the model parameters and thresholds), we mainly consider $T_{RAG}$ and $T_{ACC}$ due to they significantly impact the final performance. We show the performance under different $T_{RAG}$s in Figure \ref{fig:trag} and different $T_{ACC}$s in Figure \ref{fig:tacc}. When each threshold is considered, the other threshold is set to 0.

As for $T_{RAG}$, the trade-off between performance and latency should be considered. As depicted in Figure \ref{fig:trag}, performance exhibits a rising trend followed by a decline as $T_{RAG}$ increases. For the stage of $\textit{CARD-RG}_1$, the ES is less than invariable RG until $T_{RAG}$ reaches about 0.8. However, for $\textit{CARD-RG}_2$, ES reaches the same level when $T_{RAG}$ is only about 0.5, where the reduced ratio of RAG reaches almost 80\%. The finding is intuitively reasonable, as $RG_1$ significantly improves the performance of zero-shot generation, while $RG_2$ helps relatively little. This strengthens the necessity of our \text{\ourmethod}, which achieves better performance with much less retrieval process.

As for $T_{ACC}$, the users can inject prior knowledge of whether RAG is important to select the best result. All previous-iteration results are rejected when $T_{ACC}$ is close to 0, and all later-iteration results are preserved when $T_{ACC}$ is large enough. From Figure \ref{fig:tacc}, we see that when $T_{ACC}$ is close to 1, the ES reaches highest, implying that directly choosing results via comparing the $\hat{s}$ is effective. However, we see that the optimum $T_{ACC}$ is smaller than 1, indicating that we should trust the results of later iteration more in our scenario. However, when RAG brings more \textit{degenerated cases}, especially for the reason of overconfidence \cite{ni2024llms}, it is worth re-considering the value of $T_{ACC}$ (e.g. slightly larger than 1 to avoid more later results).

\section{Related Work}
\subsection{Repository-level Code Completion}
Automatic completing the code with repository-level context has been a challenging task~\cite{tu2014on}. Methods based on n-gram LMs~\cite{tu2014on}, recurrent neural networks~\cite{10.1145/3106237.3106290,wang2021cocosum}, and Transformers~\cite{10.1145/3368089.3417058, zhang2023repocoder} attempt this task by leveraging cross-file context. 
With the success of large LMs, researchers have applied those powerful code LMs to this task and greatly improved completion performance~\cite{wang2023codet5, rozière2024code, nijkamp2023codegen}. 
Due to the impressive in-context ability of code LMs, knowledge retrieved from other files has a significant impact on the generated results~\cite{zhang2023repocoder}. 
With retrieved contexts, large LMs show promising performance without training on the unseen repositories. However, the quality of retrieved snippets constrains the performance of RAG, and iterative RAG is experimentally demonstrated to enhance the quality of retrieved code snippets~\cite{zhang2023repocoder}, thus boosting code LMs to complete the code more precisely. 
However, multiple iterations and \textit{degenerated cases} constrain the latency and performance respectively. 
Our solution alleviates these two problems through adaptive retrieval and selective acceptance of generation results.

\subsection{Uncertainty Estimation for LMs}
The uncertainty estimation for traditional machine learning has been well studied \cite{ovadia2019trust, Abdar_2021, gawlikowski2022survey}. However, estimating the uncertainty of LMs is challenging, especially the uncertainty of sentences rather than a fixed-dimension output \cite{liu2024uncertainty}. 
For black-box LMs, measuring uncertainty can base on multiple sampled outputs, while for white-box LMs, more information can be utilized to compute the uncertainty metric more accurately. For example, some works focus on unsupervised methods via entropy \cite{malinin2021uncertainty, 10172803}, similarity \cite{fomicheva2020unsupervised, zhou-etal-2020-uncertainty, lin-etal-2022-towards} and semantic \cite{kuhn2023semantic, duan2024llms}, logits \cite{chen2024inside, kadavath2022language}, and hidden states \cite{azaria2023internal, su2024unsupervised}, calculating a metric quantifying the uncertainty. However, the benefits of supervised approaches have been increasingly noticed. For instance, the internal states and outputs show conflict in different scenarios \cite{liu2023cognitive}, underscoring the potential improvement of supervised approaches in leveraging useful information from different sources (e.g. entropy, probability, etc.). We propose a simple supervised way to estimate the uncertainty with entropy and probability (but not limited to these two) and achieve a superior estimation performance.

\subsection{Adaptive RAG}
Over retrieval can result in a waste of time and potential confusion when the model’s inherent parameterized knowledge is sufficient for answering relevant questions.
Recently, how to construct an active and adaptive RAG paradigm has attracted the attention of researchers~\cite{liang2024repofuse, zhang2023repocoder, wu2024repoformer, lu2022reacc, jiang2023active}. 
One of the core challenges of adaptive RAG is to determine \textit{when to retrieve}. Some existing methods estimate the uncertainty via token probability \cite{jiang2023active} and information entropy \cite{li-etal-2023-web} and decide whether to retrieve. 
However, without supervised training, these methods can not estimate the quality of generation quantitatively but only treat it as a classification task and provide a rough decision. 
SKR \cite{wang2023selfknowledge} employs the capability of LMs to determine whether they possess the knowledge to respond to a question. 
If they do, no retrieval is conducted. 
Self-RAG \cite{asai2023selfrag} train a critical LM to determine whether to perform retrieval. 
Similar to it, RepoFormer \cite{wu2024repoformer} trains the LM to classify whether retrieval is needed. 
It has two main limitations. First, the label would be noisy because the improvement of ES highly depends on the retrieval process. 
In other words, the label may change when different retrievers or different prompt templates are adopted. 
Second, its ability to classify is obtained via time and computation-consuming supervised fine-tuning (SFT). 
Thus, only the fine-tuned LM with the same input format as the format in the SFT stage works.
We propose a plug-in module, which provides the ability to choose \textit{when to retrieve} without training the LMs and evaluate the quality of generation.

\section{Conclusion}

In this paper, we present \text{\ourmethod}, a lightweight solution aimed at reducing unnecessary retrievals and addressing the issue of \textit{degenerated cases} encountered by previous RAG systems. 
\textit{\ourmethod} utilizes logits-based uncertainty estimation and provides two simple functions $isRetrieve$ and $Select$.
These functions can be independently employed to respectively enhance efficiency and effectiveness, serving as adaptable plugins for any existing RAG system.
Results from RepoEval demonstrate the significant reduction in latency alongside the improved completion performance across three distinct tasks achieved by our method. 
We conduct extensive experiments to assess the generalizability of \ourmethod, showcasing its efficacy across various programming languages, code LMs, and RAG systems.



\bibliographystyle{plainnat}
\bibliography{main}

\begin{thebibliography}{50}
\providecommand{\natexlab}[1]{#1}
\providecommand{\url}[1]{\texttt{#1}}
\expandafter\ifx\csname urlstyle\endcsname\relax
  \providecommand{\doi}[1]{doi: #1}\else
  \providecommand{\doi}{doi: \begingroup \urlstyle{rm}\Url}\fi

\bibitem[Abdar et~al.(2021)Abdar, Pourpanah, Hussain, Rezazadegan, Liu, Ghavamzadeh, Fieguth, Cao, Khosravi, Acharya, Makarenkov, and Nahavandi]{Abdar_2021}
Moloud Abdar, Farhad Pourpanah, Sadiq Hussain, Dana Rezazadegan, Li~Liu, Mohammad Ghavamzadeh, Paul Fieguth, Xiaochun Cao, Abbas Khosravi, U.~Rajendra Acharya, Vladimir Makarenkov, and Saeid Nahavandi.
\newblock A review of uncertainty quantification in deep learning: Techniques, applications and challenges.
\newblock \emph{Information Fusion}, 76:\penalty0 243–297, December 2021.
\newblock ISSN 1566-2535.
\newblock \doi{10.1016/j.inffus.2021.05.008}.
\newblock URL \url{http://dx.doi.org/10.1016/j.inffus.2021.05.008}.

\bibitem[Asai et~al.(2023)Asai, Wu, Wang, Sil, and Hajishirzi]{asai2023selfrag}
Akari Asai, Zeqiu Wu, Yizhong Wang, Avirup Sil, and Hannaneh Hajishirzi.
\newblock Self-rag: Learning to retrieve, generate, and critique through self-reflection, 2023.

\bibitem[Azaria and Mitchell(2023)]{azaria2023internal}
Amos Azaria and Tom Mitchell.
\newblock The internal state of an llm knows when it's lying, 2023.

\bibitem[Breiman(2001)]{breiman2001random}
Leo Breiman.
\newblock Random forests.
\newblock \emph{Machine learning}, 45:\penalty0 5--32, 2001.

\bibitem[Chen et~al.(2024)Chen, Liu, Chen, Gu, Wu, Tao, Fu, and Ye]{chen2024inside}
Chao Chen, Kai Liu, Ze~Chen, Yi~Gu, Yue Wu, Mingyuan Tao, Zhihang Fu, and Jieping Ye.
\newblock Inside: Llms' internal states retain the power of hallucination detection, 2024.

\bibitem[Chen and Guestrin(2016)]{Chen_2016}
Tianqi Chen and Carlos Guestrin.
\newblock Xgboost: A scalable tree boosting system.
\newblock In \emph{Proceedings of the 22nd ACM SIGKDD International Conference on Knowledge Discovery and Data Mining}, KDD ’16. ACM, August 2016.
\newblock \doi{10.1145/2939672.2939785}.
\newblock URL \url{http://dx.doi.org/10.1145/2939672.2939785}.

\bibitem[Cheng et~al.(2024)Cheng, Wu, and Hu]{cheng2024dataflowguided}
Wei Cheng, Yuhan Wu, and Wei Hu.
\newblock Dataflow-guided retrieval augmentation for repository-level code completion, 2024.

\bibitem[Clement et~al.(2021)Clement, Lu, Liu, Tufano, Drain, Duan, Sundaresan, and Svyatkovskiy]{clement2021longrange}
Colin~B. Clement, Shuai Lu, Xiaoyu Liu, Michele Tufano, Dawn Drain, Nan Duan, Neel Sundaresan, and Alexey Svyatkovskiy.
\newblock Long-range modeling of source code files with ewash: Extended window access by syntax hierarchy, 2021.

\bibitem[Duan et~al.(2024)Duan, Yang, and Tam]{duan2024llms}
Hanyu Duan, Yi~Yang, and Kar~Yan Tam.
\newblock Do llms know about hallucination? an empirical investigation of llm's hidden states, 2024.

\bibitem[Eghbali and Pradel(2024)]{eghbali2024dehallucinator}
Aryaz Eghbali and Michael Pradel.
\newblock De-hallucinator: Iterative grounding for llm-based code completion, 2024.

\bibitem[Fomicheva et~al.(2020)Fomicheva, Sun, Yankovskaya, Blain, Guzmán, Fishel, Aletras, Chaudhary, and Specia]{fomicheva2020unsupervised}
Marina Fomicheva, Shuo Sun, Lisa Yankovskaya, Frédéric Blain, Francisco Guzmán, Mark Fishel, Nikolaos Aletras, Vishrav Chaudhary, and Lucia Specia.
\newblock Unsupervised quality estimation for neural machine translation, 2020.

\bibitem[Gawlikowski et~al.(2022)Gawlikowski, Tassi, Ali, Lee, Humt, Feng, Kruspe, Triebel, Jung, Roscher, Shahzad, Yang, Bamler, and Zhu]{gawlikowski2022survey}
Jakob Gawlikowski, Cedrique Rovile~Njieutcheu Tassi, Mohsin Ali, Jongseok Lee, Matthias Humt, Jianxiang Feng, Anna Kruspe, Rudolph Triebel, Peter Jung, Ribana Roscher, Muhammad Shahzad, Wen Yang, Richard Bamler, and Xiao~Xiang Zhu.
\newblock A survey of uncertainty in deep neural networks, 2022.

\bibitem[Guo et~al.(2022)Guo, Lu, Duan, Wang, Zhou, and Yin]{guo2022unixcoder}
Daya Guo, Shuai Lu, Nan Duan, Yanlin Wang, Ming Zhou, and Jian Yin.
\newblock Unixcoder: Unified cross-modal pre-training for code representation, 2022.

\bibitem[Guo et~al.(2024)Guo, Zhu, Yang, Xie, Dong, Zhang, Chen, Bi, Wu, Li, Luo, Xiong, and Liang]{guo2024deepseekcoder}
Daya Guo, Qihao Zhu, Dejian Yang, Zhenda Xie, Kai Dong, Wentao Zhang, Guanting Chen, Xiao Bi, Y.~Wu, Y.~K. Li, Fuli Luo, Yingfei Xiong, and Wenfeng Liang.
\newblock Deepseek-coder: When the large language model meets programming -- the rise of code intelligence, 2024.

\bibitem[Gupta et~al.(2024)Gupta, Narasimhan, Jitkrittum, Rawat, Menon, and Kumar]{gupta2024language}
Neha Gupta, Harikrishna Narasimhan, Wittawat Jitkrittum, Ankit~Singh Rawat, Aditya~Krishna Menon, and Sanjiv Kumar.
\newblock Language model cascades: Token-level uncertainty and beyond, 2024.

\bibitem[Hellendoorn and Devanbu(2017)]{10.1145/3106237.3106290}
Vincent~J. Hellendoorn and Premkumar Devanbu.
\newblock Are deep neural networks the best choice for modeling source code?
\newblock In \emph{Proceedings of the 2017 11th Joint Meeting on Foundations of Software Engineering}, ESEC/FSE 2017, page 763–773, New York, NY, USA, 2017. Association for Computing Machinery.
\newblock ISBN 9781450351058.
\newblock \doi{10.1145/3106237.3106290}.
\newblock URL \url{https://doi.org/10.1145/3106237.3106290}.

\bibitem[Hochreiter and Schmidhuber(1997)]{6795963}
Sepp Hochreiter and Jürgen Schmidhuber.
\newblock Long short-term memory.
\newblock \emph{Neural Computation}, 9\penalty0 (8):\penalty0 1735--1780, 1997.
\newblock \doi{10.1162/neco.1997.9.8.1735}.

\bibitem[Jiang et~al.(2023)Jiang, Xu, Gao, Sun, Liu, Dwivedi-Yu, Yang, Callan, and Neubig]{jiang2023active}
Zhengbao Jiang, Frank~F. Xu, Luyu Gao, Zhiqing Sun, Qian Liu, Jane Dwivedi-Yu, Yiming Yang, Jamie Callan, and Graham Neubig.
\newblock Active retrieval augmented generation, 2023.

\bibitem[Kadavath et~al.(2022)Kadavath, Conerly, Askell, Henighan, Drain, Perez, Schiefer, Hatfield-Dodds, DasSarma, Tran-Johnson, Johnston, El-Showk, Jones, Elhage, Hume, Chen, Bai, Bowman, Fort, Ganguli, Hernandez, Jacobson, Kernion, Kravec, Lovitt, Ndousse, Olsson, Ringer, Amodei, Brown, Clark, Joseph, Mann, McCandlish, Olah, and Kaplan]{kadavath2022language}
Saurav Kadavath, Tom Conerly, Amanda Askell, Tom Henighan, Dawn Drain, Ethan Perez, Nicholas Schiefer, Zac Hatfield-Dodds, Nova DasSarma, Eli Tran-Johnson, Scott Johnston, Sheer El-Showk, Andy Jones, Nelson Elhage, Tristan Hume, Anna Chen, Yuntao Bai, Sam Bowman, Stanislav Fort, Deep Ganguli, Danny Hernandez, Josh Jacobson, Jackson Kernion, Shauna Kravec, Liane Lovitt, Kamal Ndousse, Catherine Olsson, Sam Ringer, Dario Amodei, Tom Brown, Jack Clark, Nicholas Joseph, Ben Mann, Sam McCandlish, Chris Olah, and Jared Kaplan.
\newblock Language models (mostly) know what they know, 2022.

\bibitem[Ke et~al.(2017)Ke, Meng, Finley, Wang, Chen, Ma, Ye, and Liu]{NIPS2017_6449f44a}
Guolin Ke, Qi~Meng, Thomas Finley, Taifeng Wang, Wei Chen, Weidong Ma, Qiwei Ye, and Tie-Yan Liu.
\newblock Lightgbm: A highly efficient gradient boosting decision tree.
\newblock In I.~Guyon, U.~Von Luxburg, S.~Bengio, H.~Wallach, R.~Fergus, S.~Vishwanathan, and R.~Garnett, editors, \emph{Advances in Neural Information Processing Systems}, volume~30. Curran Associates, Inc., 2017.
\newblock URL \url{https://proceedings.neurips.cc/paper_files/paper/2017/file/6449f44a102fde848669bdd9eb6b76fa-Paper.pdf}.

\bibitem[Kocetkov et~al.(2022)Kocetkov, Li, Allal, Li, Mou, Ferrandis, Jernite, Mitchell, Hughes, Wolf, Bahdanau, von Werra, and de~Vries]{kocetkov2022stack}
Denis Kocetkov, Raymond Li, Loubna~Ben Allal, Jia Li, Chenghao Mou, Carlos~Muñoz Ferrandis, Yacine Jernite, Margaret Mitchell, Sean Hughes, Thomas Wolf, Dzmitry Bahdanau, Leandro von Werra, and Harm de~Vries.
\newblock The stack: 3 tb of permissively licensed source code, 2022.

\bibitem[Kuhn et~al.(2023)Kuhn, Gal, and Farquhar]{kuhn2023semantic}
Lorenz Kuhn, Yarin Gal, and Sebastian Farquhar.
\newblock Semantic uncertainty: Linguistic invariances for uncertainty estimation in natural language generation, 2023.

\bibitem[Kwon et~al.(2023)Kwon, Li, Zhuang, Sheng, Zheng, Yu, Gonzalez, Zhang, and Stoica]{kwon2023efficient}
Woosuk Kwon, Zhuohan Li, Siyuan Zhuang, Ying Sheng, Lianmin Zheng, Cody~Hao Yu, Joseph~E. Gonzalez, Hao Zhang, and Ion Stoica.
\newblock Efficient memory management for large language model serving with pagedattention, 2023.

\bibitem[Levenshtein(1965)]{Levenshtein1965BinaryCC}
Vladimir~I. Levenshtein.
\newblock Binary codes capable of correcting deletions, insertions, and reversals.
\newblock \emph{Soviet physics. Doklady}, 10:\penalty0 707--710, 1965.
\newblock URL \url{https://api.semanticscholar.org/CorpusID:60827152}.

\bibitem[Li et~al.(2023)Li, Tang, Zhao, Wang, Nie, and Wen]{li-etal-2023-web}
Junyi Li, Tianyi Tang, Wayne~Xin Zhao, Jingyuan Wang, Jian-Yun Nie, and Ji-Rong Wen.
\newblock The web can be your oyster for improving language models.
\newblock In Anna Rogers, Jordan Boyd-Graber, and Naoaki Okazaki, editors, \emph{Findings of the Association for Computational Linguistics: ACL 2023}, pages 728--746, Toronto, Canada, July 2023. Association for Computational Linguistics.
\newblock \doi{10.18653/v1/2023.findings-acl.46}.
\newblock URL \url{https://aclanthology.org/2023.findings-acl.46}.

\bibitem[Liang et~al.(2024)Liang, Xie, Zhang, Zheng, Di, wei jiang, Chen, Wang, and Fan]{liang2024repofuse}
Ming Liang, Xiaoheng Xie, Gehao Zhang, Xunjin Zheng, Peng Di, wei jiang, Hongwei Chen, Chengpeng Wang, and Gang Fan.
\newblock Repofuse: Repository-level code completion with fused dual context, 2024.

\bibitem[Lin et~al.(2022)Lin, Liu, and Shang]{lin-etal-2022-towards}
Zi~Lin, Jeremiah~Zhe Liu, and Jingbo Shang.
\newblock Towards collaborative neural-symbolic graph semantic parsing via uncertainty.
\newblock In Smaranda Muresan, Preslav Nakov, and Aline Villavicencio, editors, \emph{Findings of the Association for Computational Linguistics: ACL 2022}, pages 4160--4173, Dublin, Ireland, May 2022. Association for Computational Linguistics.
\newblock \doi{10.18653/v1/2022.findings-acl.328}.
\newblock URL \url{https://aclanthology.org/2022.findings-acl.328}.

\bibitem[Liu et~al.(2023)Liu, Casper, Hadfield-Menell, and Andreas]{liu2023cognitive}
Kevin Liu, Stephen Casper, Dylan Hadfield-Menell, and Jacob Andreas.
\newblock Cognitive dissonance: Why do language model outputs disagree with internal representations of truthfulness?, 2023.

\bibitem[Liu et~al.(2024)Liu, Pan, Li, and Chen]{liu2024uncertainty}
Linyu Liu, Yu~Pan, Xiaocheng Li, and Guanting Chen.
\newblock Uncertainty estimation and quantification for llms: A simple supervised approach.
\newblock \emph{arXiv preprint arXiv:2404.15993}, 2024.

\bibitem[Lozhkov et~al.(2024)Lozhkov, Li, Allal, Cassano, Lamy-Poirier, Tazi, Tang, Pykhtar, Liu, Wei, Liu, Tian, Kocetkov, Zucker, Belkada, Wang, Liu, Abulkhanov, Paul, Li, Li, Risdal, Li, Zhu, Zhuo, Zheltonozhskii, Dade, Yu, Krauß, Jain, Su, He, Dey, Abati, Chai, Muennighoff, Tang, Oblokulov, Akiki, Marone, Mou, Mishra, Gu, Hui, Dao, Zebaze, Dehaene, Patry, Xu, McAuley, Hu, Scholak, Paquet, Robinson, Anderson, Chapados, Patwary, Tajbakhsh, Jernite, Ferrandis, Zhang, Hughes, Wolf, Guha, von Werra, and de~Vries]{lozhkov2024starcoder}
Anton Lozhkov, Raymond Li, Loubna~Ben Allal, Federico Cassano, Joel Lamy-Poirier, Nouamane Tazi, Ao~Tang, Dmytro Pykhtar, Jiawei Liu, Yuxiang Wei, Tianyang Liu, Max Tian, Denis Kocetkov, Arthur Zucker, Younes Belkada, Zijian Wang, Qian Liu, Dmitry Abulkhanov, Indraneil Paul, Zhuang Li, Wen-Ding Li, Megan Risdal, Jia Li, Jian Zhu, Terry~Yue Zhuo, Evgenii Zheltonozhskii, Nii Osae~Osae Dade, Wenhao Yu, Lucas Krauß, Naman Jain, Yixuan Su, Xuanli He, Manan Dey, Edoardo Abati, Yekun Chai, Niklas Muennighoff, Xiangru Tang, Muhtasham Oblokulov, Christopher Akiki, Marc Marone, Chenghao Mou, Mayank Mishra, Alex Gu, Binyuan Hui, Tri Dao, Armel Zebaze, Olivier Dehaene, Nicolas Patry, Canwen Xu, Julian McAuley, Han Hu, Torsten Scholak, Sebastien Paquet, Jennifer Robinson, Carolyn~Jane Anderson, Nicolas Chapados, Mostofa Patwary, Nima Tajbakhsh, Yacine Jernite, Carlos~Muñoz Ferrandis, Lingming Zhang, Sean Hughes, Thomas Wolf, Arjun Guha, Leandro von Werra, and Harm de~Vries.
\newblock Starcoder 2 and the stack v2: The next generation, 2024.

\bibitem[Lu et~al.(2022)Lu, Duan, Han, Guo, won Hwang, and Svyatkovskiy]{lu2022reacc}
Shuai Lu, Nan Duan, Hojae Han, Daya Guo, Seung won Hwang, and Alexey Svyatkovskiy.
\newblock Reacc: A retrieval-augmented code completion framework, 2022.

\bibitem[MacQueen(1967)]{MacQueen1967SomeMF}
J.~MacQueen.
\newblock Some methods for classification and analysis of multivariate observations.
\newblock 1967.
\newblock URL \url{https://api.semanticscholar.org/CorpusID:6278891}.

\bibitem[Malinin and Gales(2021)]{malinin2021uncertainty}
Andrey Malinin and Mark Gales.
\newblock Uncertainty estimation in autoregressive structured prediction, 2021.

\bibitem[Ni et~al.(2024)Ni, Bi, Guo, and Cheng]{ni2024llms}
Shiyu Ni, Keping Bi, Jiafeng Guo, and Xueqi Cheng.
\newblock When do llms need retrieval augmentation? mitigating llms' overconfidence helps retrieval augmentation, 2024.

\bibitem[Nijkamp et~al.(2023)Nijkamp, Pang, Hayashi, Tu, Wang, Zhou, Savarese, and Xiong]{nijkamp2023codegen}
Erik Nijkamp, Bo~Pang, Hiroaki Hayashi, Lifu Tu, Huan Wang, Yingbo Zhou, Silvio Savarese, and Caiming Xiong.
\newblock Codegen: An open large language model for code with multi-turn program synthesis, 2023.

\bibitem[Ovadia et~al.(2019)Ovadia, Fertig, Ren, Nado, Sculley, Nowozin, Dillon, Lakshminarayanan, and Snoek]{ovadia2019trust}
Yaniv Ovadia, Emily Fertig, Jie Ren, Zachary Nado, D~Sculley, Sebastian Nowozin, Joshua~V. Dillon, Balaji Lakshminarayanan, and Jasper Snoek.
\newblock Can you trust your model's uncertainty? evaluating predictive uncertainty under dataset shift, 2019.

\bibitem[Rozière et~al.(2024)Rozière, Gehring, Gloeckle, Sootla, Gat, Tan, Adi, Liu, Sauvestre, Remez, Rapin, Kozhevnikov, Evtimov, Bitton, Bhatt, Ferrer, Grattafiori, Xiong, Défossez, Copet, Azhar, Touvron, Martin, Usunier, Scialom, and Synnaeve]{rozière2024code}
Baptiste Rozière, Jonas Gehring, Fabian Gloeckle, Sten Sootla, Itai Gat, Xiaoqing~Ellen Tan, Yossi Adi, Jingyu Liu, Romain Sauvestre, Tal Remez, Jérémy Rapin, Artyom Kozhevnikov, Ivan Evtimov, Joanna Bitton, Manish Bhatt, Cristian~Canton Ferrer, Aaron Grattafiori, Wenhan Xiong, Alexandre Défossez, Jade Copet, Faisal Azhar, Hugo Touvron, Louis Martin, Nicolas Usunier, Thomas Scialom, and Gabriel Synnaeve.
\newblock Code llama: Open foundation models for code, 2024.

\bibitem[Shao et~al.(2023)Shao, Gong, Shen, Huang, Duan, and Chen]{shao2023enhancing}
Zhihong Shao, Yeyun Gong, Yelong Shen, Minlie Huang, Nan Duan, and Weizhu Chen.
\newblock Enhancing retrieval-augmented large language models with iterative retrieval-generation synergy, 2023.

\bibitem[Su et~al.(2024)Su, Wang, Ai, HU, Wu, Zhou, and Liu]{su2024unsupervised}
Weihang Su, Changyue Wang, Qingyao Ai, Yiran HU, Zhijing Wu, Yujia Zhou, and Yiqun Liu.
\newblock Unsupervised real-time hallucination detection based on the internal states of large language models, 2024.

\bibitem[Svyatkovskiy et~al.(2020)Svyatkovskiy, Deng, Fu, and Sundaresan]{10.1145/3368089.3417058}
Alexey Svyatkovskiy, Shao~Kun Deng, Shengyu Fu, and Neel Sundaresan.
\newblock Intellicode compose: code generation using transformer.
\newblock In \emph{Proceedings of the 28th ACM Joint Meeting on European Software Engineering Conference and Symposium on the Foundations of Software Engineering}, ESEC/FSE 2020, page 1433–1443, New York, NY, USA, 2020. Association for Computing Machinery.
\newblock ISBN 9781450370431.
\newblock \doi{10.1145/3368089.3417058}.
\newblock URL \url{https://doi.org/10.1145/3368089.3417058}.

\bibitem[Tan et~al.(2024)Tan, Luo, Jiang, Zhan, Li, Zhang, and Zhang]{tan2024promptbased}
Hanzhuo Tan, Qi~Luo, Ling Jiang, Zizheng Zhan, Jing Li, Haotian Zhang, and Yuqun Zhang.
\newblock Prompt-based code completion via multi-retrieval augmented generation, 2024.

\bibitem[Tu et~al.(2014)Tu, Su, and Devanbu]{tu2014on}
Zhaopeng Tu, Zhendong Su, and Premkumar Devanbu.
\newblock On the localness of software.
\newblock In \emph{Proceedings of the 22nd ACM SIGSOFT International Symposium on Foundations of Software Engineering}, FSE 2014, page 269–280, New York, NY, USA, 2014. Association for Computing Machinery.
\newblock ISBN 9781450330565.
\newblock \doi{10.1145/2635868.2635875}.
\newblock URL \url{https://doi.org/10.1145/2635868.2635875}.

\bibitem[Wang et~al.(2021)Wang, Shi, Du, Yang, Hu, Han, Zhang, and Zhang]{wang2021cocosum}
Yanlin Wang, Ensheng Shi, Lun Du, Xiaodi Yang, Yuxuan Hu, Shi Han, Hongyu Zhang, and Dongmei Zhang.
\newblock Cocosum: Contextual code summarization with multi-relational graph neural network, 2021.

\bibitem[Wang et~al.(2023{\natexlab{a}})Wang, Li, Sun, and Liu]{wang2023selfknowledge}
Yile Wang, Peng Li, Maosong Sun, and Yang Liu.
\newblock Self-knowledge guided retrieval augmentation for large language models, 2023{\natexlab{a}}.

\bibitem[Wang et~al.(2023{\natexlab{b}})Wang, Le, Gotmare, Bui, Li, and Hoi]{wang2023codet5}
Yue Wang, Hung Le, Akhilesh~Deepak Gotmare, Nghi D.~Q. Bui, Junnan Li, and Steven C.~H. Hoi.
\newblock Codet5+: Open code large language models for code understanding and generation, 2023{\natexlab{b}}.

\bibitem[Wu et~al.(2024)Wu, Ahmad, Zhang, Ramanathan, and Ma]{wu2024repoformer}
Di~Wu, Wasi~Uddin Ahmad, Dejiao Zhang, Murali~Krishna Ramanathan, and Xiaofei Ma.
\newblock Repoformer: Selective retrieval for repository-level code completion, 2024.

\bibitem[Xia et~al.(2023)Xia, Wei, and Zhang]{10172803}
Chunqiu~Steven Xia, Yuxiang Wei, and Lingming Zhang.
\newblock Automated program repair in the era of large pre-trained language models.
\newblock In \emph{2023 IEEE/ACM 45th International Conference on Software Engineering (ICSE)}, pages 1482--1494, 2023.
\newblock \doi{10.1109/ICSE48619.2023.00129}.

\bibitem[Zhang* et~al.(2024)Zhang*, Ahmad*, Tan, Ding, Nallapati, Roth, Ma, and Xiang]{zhang2024codesage}
Dejiao Zhang*, Wasi Ahmad*, Ming Tan, Hantian Ding, Ramesh Nallapati, Dan Roth, Xiaofei Ma, and Bing Xiang.
\newblock Codesage: Code representation learning at scale.
\newblock In \emph{The Twelfth International Conference on Learning Representations}, 2024.
\newblock URL \url{https://openreview.net/forum?id=vfzRRjumpX}.

\bibitem[Zhang et~al.(2023)Zhang, Chen, Zhang, Keung, Liu, Zan, Mao, Lou, and Chen]{zhang2023repocoder}
Fengji Zhang, Bei Chen, Yue Zhang, Jacky Keung, Jin Liu, Daoguang Zan, Yi~Mao, Jian-Guang Lou, and Weizhu Chen.
\newblock Repocoder: Repository-level code completion through iterative retrieval and generation, 2023.

\bibitem[Zhou et~al.(2020)Zhou, Yang, Wong, Wan, and Chao]{zhou-etal-2020-uncertainty}
Yikai Zhou, Baosong Yang, Derek~F. Wong, Yu~Wan, and Lidia~S. Chao.
\newblock Uncertainty-aware curriculum learning for neural machine translation.
\newblock In Dan Jurafsky, Joyce Chai, Natalie Schluter, and Joel Tetreault, editors, \emph{Proceedings of the 58th Annual Meeting of the Association for Computational Linguistics}, pages 6934--6944, Online, July 2020. Association for Computational Linguistics.
\newblock \doi{10.18653/v1/2020.acl-main.620}.
\newblock URL \url{https://aclanthology.org/2020.acl-main.620}.

\end{thebibliography}
\end{document}